\documentclass[twocolumn]{aastex62}

\usepackage{threeparttable}
\usepackage{rotating}
\usepackage{graphicx}

\usepackage{lineno}

\submitjournal{ApJ}

\begin{document}

\title{Marginally stable nuclear burning triggered at different depths of the neutron star surface in Low-mass X-ray binary 4U 1608--52}

\correspondingauthor{}
\email{lvming@xtu.edu.cn}

\author{Ming Lyu}
\affiliation{Department of Physics, Xiangtan University, Xiangtan, Hunan 411105, China}
\affiliation{Key Laboratory of Stars and Interstellar Medium, Xiangtan University, Xiangtan, Hunan 411105, China}

\author{Guobao Zhang}
\affiliation{Yunnan Observatories, Chinese Academy of Sciences (CAS), Kunming 650216, P.R. China}
\affiliation{Key Laboratory for the Structure and Evolution of Celestial Objects, CAS, Kunming 650216, P.R. China}

\author{Mariano M\'endez}
\affiliation{Kapteyn Astronomical Institute, University of Groningen, PO BOX 800, NL-9700 AV Groningen, the Netherlands}

\author{Huaping Xiao}
\affiliation{Department of Physics, Xiangtan University, Xiangtan, Hunan 411105, China}
\affiliation{Key Laboratory of Stars and Interstellar Medium, Xiangtan University, Xiangtan, Hunan 411105, China}




\begin{abstract}
We investigated the timing and spectral properties of the millihertz quasi-periodic oscillations (mHz QPOs) in the neutron-star low-mass X-ray binary 4U 1608--52 using NICER observations. Our analysis reveals a correlation between the QPO frequency and its absolute amplitude, as well as between the frequency and the temperature of the burning layer. Intensity-resolved spectral analysis indicates that the flux modulation of the mHz QPOs is primarily caused by the variations in the blackbody temperature in most observations. Furthermore, for the first time, we report that as the source evolves from the soft spectral state toward the transitional state, the marginally stable burning responsible for the mHz QPOs ignites at deeper layers of the neutron-star surface. The radiation flux associated with the mHz QPOs shows a decreasing trend as the source moves into the transitional state. These two findings support a scenario in which the marginally stable nuclear burning ignites at deeper layers as the temperature decreases, releasing less energy from the nuclear reaction. Finally, we determine that the energy release rate of the marginally stable burning is around $\sim$ 10$^{35}$ erg/s, consistent with the theoretical predictions.

\end{abstract}



\section{Introduction}

Millihertz quasi-periodic oscillations (mHz QPOs) in neutron-star (NS) low-mass X-ray binaries (LMXBs) were first detected by \citet{revni01} using Rossi X-ray Timing Explorer (RXTE) observations. These QPOs typically oscillate at very low frequencies, approximately $\sim$ 6-14 mHz \citep{revni01,diego08,lyu15}. Moreover, the mHz QPOs are detected only within a certain source luminosity range and disappear or become undetectable above 5 keV.

The mHz QPOs are closely connected to the type I X-ray thermal nuclear bursts in observations. \citet{revni01} showed that the mHz QPOs disappeared when there was a X-ray burst, and this phenomenon was further confirmed by subsequent observations \citep{diego08,lyu15,lyu16,Mancuso19}. Moreover, \citet{diego08} found that the frequency of the mHz QPOs systematically decreases before X-ray bursts. Later, \citet{Mancuso19} observed a similar frequency drift behavior in the LMXB EXO 0748--676. A more direct connection between the mHz QPOs and the type I X-ray bursts was reported by \citet{linaries12} in the neutron-star transient source IGR J17480--2446: as the source luminosity increased, type I X-ray bursts gradually evolved into a mHz QPO, and vice versa.

It has been proposed that mHz QPOs originate from an oscillatory mode of nuclear burning on the neutron-star surface \citep{revni01,heger07,keek09,keek14}. \citet{heger07} found that marginally stable nuclear burning of helium on the neutron-star surface could generate oscillations on timescales of $\sim$ 100 s, consistent with the $\sim$ 2 minutes period of the mHz QPOs in observations. Their model also indicates that the oscillations should only appear within a very narrow range of X-ray luminosity. However, in that model, the mHz QPOs occur at an accretion rates close to the Eddington limit, roughly an order of magnitude higher than the global accretion rate inferred from the observed X-ray luminosity when the QPOs are present.

To explain the discrepancy between the observations and the models regarding the accretion rate at which the mHz QPOs occur, \citet{heger07} proposed that the local accretion rate in the burning layer could be higher than the global rate across the entire neutron-star surface, and that mHz QPOs are therefore triggered by the local accretion rate of the burning region. Subsequently, \citet{lyu16} examined the convexity during the rising phase of Type I X-ray bursts associated with the mHz QPOs. They found that these bursts all show positive convexity and short rising time, suggesting that the mHz QPOs originate near the neutron-star equator. This result supports the theory that it is the local accretion rate that triggers the mHz QPOs: accreted material first impacts the equatorial region and then spreads over the entire neutron star surface, so the local accretion rate at the equator is high enough to drive the marginally stable burning responsible for the mHz QPOs. Whereas at higher latitudes the local accretion rate remains too low to trigger such oscillations. 

\citet{keek09} proposed an alternative scenario to explain the low accretion rate at which the mHz QPOs are observed. In their model, during the rotationally induced transport processes, the fuel becomes fully mixed to greater depths; if there is a higher heat flux from the crust, then mHz QPOs can be triggered at the observed accretion rate. Additionally, \citet{keek14} found that no plausible variations in the fuel composition and nuclear reaction rates could generate the mHz QPOs at the observed accretion rates.

Several observational properties of the mHz QPOs have already been explored in previous studies. \citet{stiele16} investigated the phase-resolved energy spectra of the mHz QPOs in 4U 1636--53 and suggested that the oscillations are driven by the modulation of the emitting area. In contrast, \citet{strohmayer18} found that the mHz QPOs in the "clocked burster" GS 1826--238 are caused by variations in the blackbody temperature. \citet{lyu19} systematically studied the timing properties of the QPOs in 4U 1636--53 using a large number of RXTE observations. Later, \citet{lyu20} reported that the fractional rms amplitude of the mHz QPOs in 4U 1636--53 increases from $\sim$ 0.2 keV to $\sim$ 3 keV, differing from the decreasing trend that has been previously reported above $\sim$ 3 keV. \citet{Mancuso21} described a downward frequency drift of the mHz QPOs in the LMXBs 4U 1608--52 and Aql X-1. \citet{fei21} studied the harmonics of the mHz QPOs using RXTE data and found that not all observations with the mHz QPOs exhibit the harmonic component. \citet{hsieh20} concluded that the mHz QPOs in 4U 1636--53 are primarily due to the modulation of the area emitting blackbody radiation. Most recently, \citet{xiao25} reported that there is a correlation between mHz QPOs and temperature of Comptonization seed photons from the neutron star surface.

In this work, we investigate the timing and spectral properties of the mHz QPOs in LMXB 4U 1608--52 using NICER observations. We focus on the intensity-resolved spectra to explore the physical parameters associated with the marginally stable nuclear burning process. We derive the column depth of the burning region and the radiation flux linked to the mHz QPOs in this source. The paper is structured as follows: The details of the data reduction and the analysis is shown in Section 2 and 3, and the corresponding results are presented in Section 4. In Section 5, we discuss our findings in the context of current marginally stable nuclear burning models.

\section{Observations and data reduction}

\begin{figure}
\center
\includegraphics[width=0.5\textwidth]{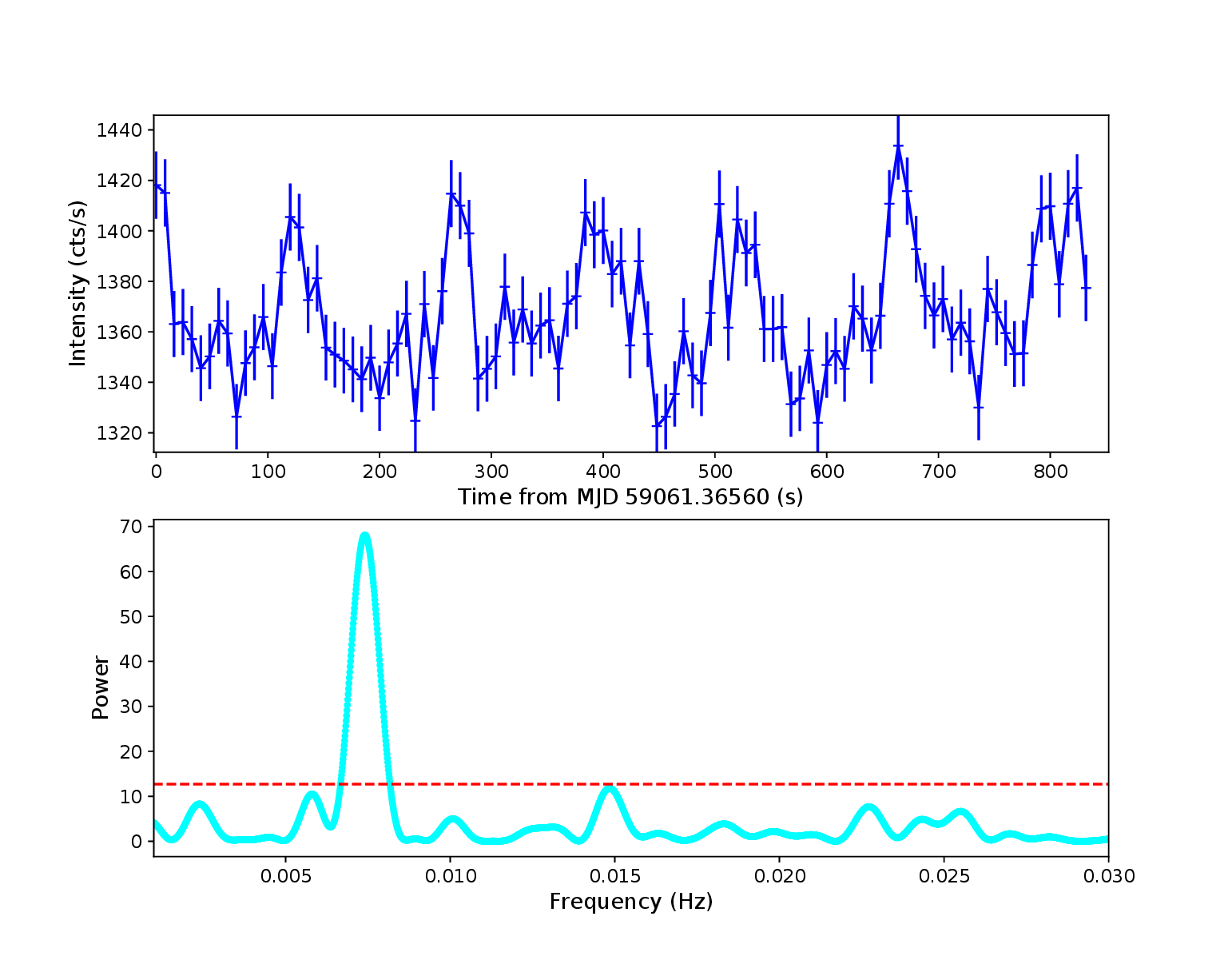}
\caption{The light curve (top) and the Lomb-Scargle periodogram (bottom) of the data set spanning the time interval 0-839 s in Obs 3657025701 with the mHz QPO. We oversampled the frequency by a factor of 100, and marked the 3$\sigma$ significance detection level with a red horizontal dashed line. }
\label{LS}
\end{figure}

\begin{figure}
\center
\includegraphics[width=0.48\textwidth]{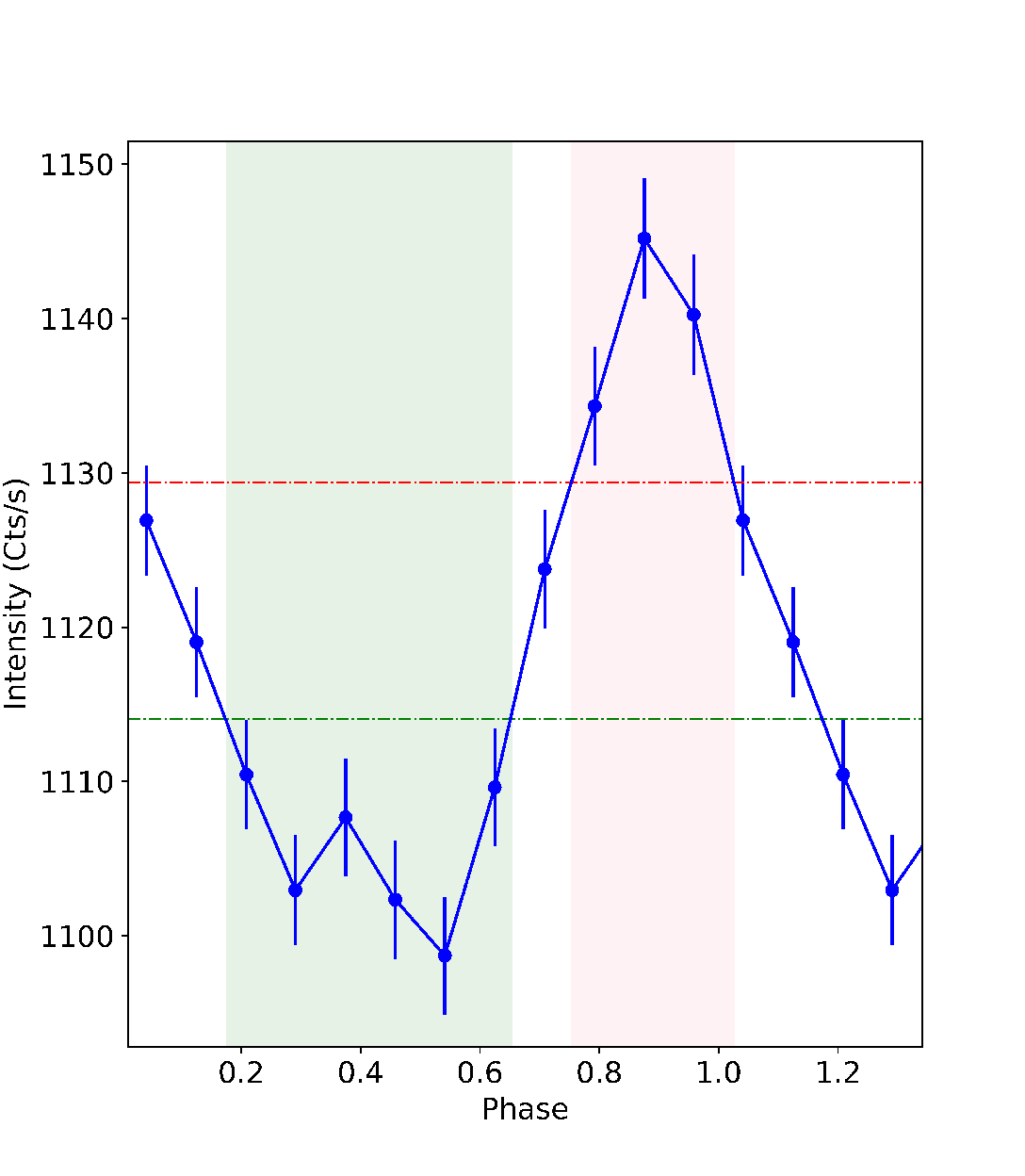}
\caption{Folded mHz QPO profile in the time interval 22238-23194 s of observation 0050070103. The red and green dashed horizontal lines mark the 2/3 and 1/3 peak photon flux level, which divide the profile into the peak stage (red shadow region), the bottom stage (green shadow region) and the middle stage (the rest). We plot $\sim$1.4 cycles to show the shape of the QPO profile clearly in the plot. }
\label{fold}
\end{figure}

\begin{table}
\centering
\caption{The mHz QPOs in the NICER observations of 4U 1608--52. Errors in this table are at the 68\% confidence level.}
\scalebox{0.65}{
\begin{tabular}{lccc}
\hline
 ObsID   & QPO epoch (s)  & Freq (mHz) & rms amplitude (\%)    \\
\hline
 0050070102  &   16679-17652   & 7.14$\pm$0.03 &   1.32$\pm$0.09   \\    
             &   22261-23195   & 6.61$\pm$0.02 &   1.21$\pm$0.09   \\         
             &   27782-28769   & 7.59$\pm$0.02 &   1.22$\pm$0.09   \\         
             &   55624-56555   & 6.68$\pm$0.06 &   0.80$\pm$0.09   \\         
             &   61143-61580   & 9.02$\pm$0.09 &   1.16$\pm$0.14   \\         
 0050070103  &   16680-17636   & 8.12$\pm$0.04 &   0.97$\pm$0.10   \\         
             &   22238-23194   & 6.61$\pm$0.04 &   1.30$\pm$0.10   \\         
 1050070101  &   0-771         & 3.71$\pm$0.06 &   2.69$\pm$0.26   \\         
             &   11121-11887   & 3.69$\pm$0.06 &   1.89$\pm$0.26   \\         
             &   22196-23002   & 3.29$\pm$0.06 &   2.95$\pm$0.27   \\         
 1050070128  &   0-861         & 6.14$\pm$0.05 &   0.75$\pm$0.08   \\         
             &   28000-28693   & 8.83$\pm$0.05 &   1.30$\pm$0.10   \\         
 1050070129  &   0-871         & 7.31$\pm$0.02 &   1.68$\pm$0.10   \\         
 1050070130  &   61178-61754   & 5.46$\pm$0.06 &   1.70$\pm$0.14   \\         
 3657020301  &   38907-39466   & 6.29$\pm$0.07 &   0.73$\pm$0.08   \\         
 3657022001  &   27879-28669   & 7.26$\pm$0.05 &   0.52$\pm$0.07   \\         
             &   33458-34189   & 8.36$\pm$0.04 &   0.67$\pm$0.07   \\         
 3657025701  &   0-839         & 7.45$\pm$0.03 &   1.28$\pm$0.09   \\         
             &   11264-12019   & 5.96$\pm$0.06 &   1.01$\pm$0.10   \\         
             &   33460-34359   & 7.38$\pm$0.04 &   0.84$\pm$0.09   \\         
             &   44619-45619   & 7.62$\pm$0.03 &   1.29$\pm$0.09   \\         
 3657025801  &   17229-17811   & 7.03$\pm$0.07 &   1.37$\pm$0.11   \\         
 3657026201  &   22231-22691   & 6.82$\pm$0.09 &   1.21$\pm$0.14   \\         
             &   33409-33831   & 7.32$\pm$0.13 &   1.01$\pm$0.14   \\         
 3657026401  &   0-769         & 5.82$\pm$0.05 &   1.51$\pm$0.11   \\         
 3657026601  &   0-724     & 4.18$\pm$0.04 &   1.49$\pm$0.13   \\         
 3657026901  &   0-1649     & 4.74$\pm$0.02 &   1.44$\pm$0.11   \\         
 8050070103  &   10900-12100   & 5.93$\pm$0.04 &   0.80$\pm$0.08   \\    

\hline          
\end{tabular}   
}
\medskip        
\\                  
\label{obs}     
\end{table} 

\begin{figure*}
\center
\includegraphics[width=1.0\textwidth]{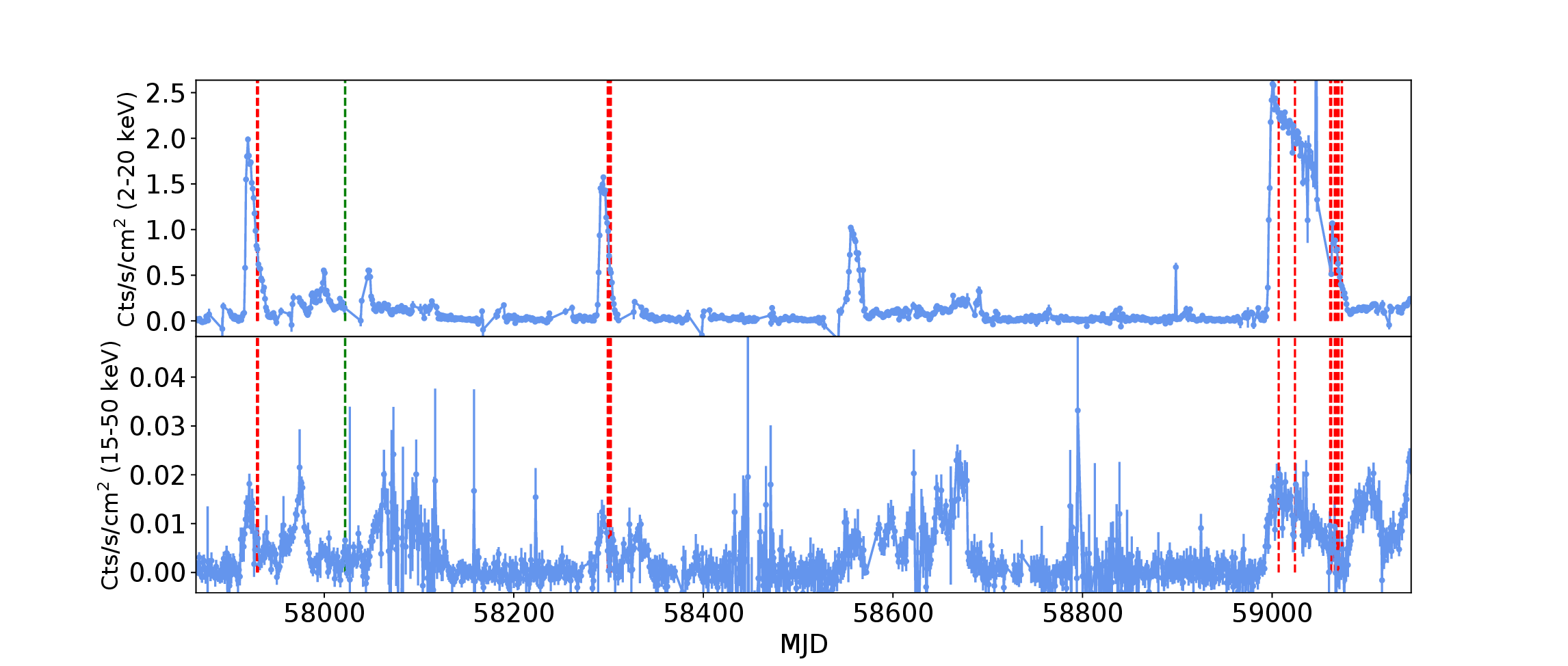}
\caption{Long-term 1-day MAXI (top) and Swift-BAT (bottom) light curve of 4U 1608--52. We mark the NICER observations with mHz QPOs in the figure: most of them are indicated with vertical red dashed lines, while Obs 1050070101 (MJD $\sim$ 58022) is marked with a green line, as it is the only detection occurring outside outbursts. For clarity, Obs 8050070103 (MJD $\sim$ 60756) is not shown.}
\label{maxi_swift}
\end{figure*}

We analyzed all archival NICER observations with exposure time larger than 800 s from June 24, 2017 to March 22, 2025 using the NICER Data Analysis Software in HEASOFT ver 6.34. NICER collects photons in an energy range of $0.2-12$ keV, with a collecting effective area more than 2000 cm$^{2}$ at 1.5 keV and $\sim$ 600 cm$^{2}$ at 6 keV. NICER has an energy solution of 85 eV at $\sim$ 1 keV and a time resolution of 40 nsec. We used the full level 2 calibration and screening pipeline {\tt nicerl2} to clean and screen the raw data with the latest calibration files (20240206). We then extracted a 1-s time resolution light curve in the 0.2-5.0 keV band for each observation using {\tt xselect} and searched potential mHz QPOs in every 1000 s segment with Lomb-Scargle (LS) periodogram \citep{lomb76,scargle82}. We identified 28 datasets that show significant mHz QPOs above the $3\sigma$ confidence level in the LS diagram (see Figure \ref{LS} for an example). Furthermore, as shown in the light curve in the top panel of Figure \ref{LS}, possible harmonic components likely exist in the dataset, similar to those found in another mHz QPO source 4U 1636--53 \citep{fei21}. To accurately derive the properties of the QPOs, we fitted the light curves of those data sets using a model $L(t)=C + \Sigma_{k=1}^{3} A_{k} sin(2\pi kft + \phi_{k})$, in which the three sine functions describe the QPO and its possible first and second harmonics, respectively. We then calculated the fractional rms amplitude of the mHz QPOs (Table \ref{obs}) using the formula, $rms=A_{1}/[\sqrt{2} \times C]$, where $A_{1}$ is the amplitude of the first sine function and C is the value of the constant component. To investigate the distribution of the mHz QPOs across different spectral states, we generated a hardness-Intensity diagram (HID) of 4U 1608-52 using the NICER observations. Hardness is defined as the count ratio between 4-10 keV and 0.5-4 keV, and the intensity is defined as the counts from 0.5 to 10 keV. 

We used {\tt nicerl3-spec} to extract the total spectra from all the mHz QPO epochs for each observation. The {\tt nicerl3-spec} is a complete end-to-end spectral product pipeline that generates a spectrum, background and associated responses following NICER's recommended procedures. We selected the SCORPEON background model in the format of a ''file'' (bkgformat=file), and used the {\tt grppha} to group the spectra with a minimum of 25 counts per bin.  

To explore the spectral properties across different QPO stages, we folded each dataset at its corresponding QPO period to obtain the folded QPO profile. The entire profile was segmented into three parts based on photon flux levels: a peak region (above 2/3 of the peak flux), a middle region (between 1/3 and 2/3 of the peak flux), and a bottom region (below 1/3 of the peak flux), as illustrated in Figure \ref{fold}. We then generated a total GTI file comprising all the peak regions from the different datasets of the same observation and extracted a combined peak spectrum for that observation. The same procedure was applied to generate the combined middle and bottom spectra. Additionally, we extracted two further intensity-resolved spectra per observation: a stable-flux spectrum (S), comprising intervals with flux below 15\% of the peak, and a stable-flux plus QPO-flux spectrum (S+Q), defined as intervals where the flux exceeded 85\% of the peak photon flux. Roughly speaking, the S spectra represents the stable flux, while the S+Q spectra corresponds to the stable flux plus the oscillatory flux. Therefore, the difference between the S and S+Q spectra is close to the oscillational flux coming from the QPOs. \footnote{We also extracted the S and S+Q spectra using 1/3 and 2/3 peak photon flux thresholds and obtained similar fitting results, indicating that the conclusions in this work are insensitive to the exact choice of the flux threshold.}

\section{Spectral analysis}
We used PyXspec in Heasoft 6.34 to fit the NICER spectra in this work in 0.4-10.0 keV energy range. To avoid possible degeneracy problems due to the limited number of source photons, here we selected only those observations with a total exposure time of the mHz QPO epochs exceeding 1500 s for the subsequent spectral analysis. We first modeled the total spectra extracted from all mHz QPO epochs using the model {\tt bbodyrad+comptt}. The {\tt bbodyrad} component describes the thermal radiation from the neutron star surface, and the {\tt comptt} component \citep{Titarchuk94} is an analytic model describing Comptonization of soft photons in a hot plasma. The model parameters are the blackbody temperature, $kT_{BB}$, the normalization, $Nor_{BB}$, the redshift, $z$, the input seed photon temperature, $kT_{seed}$, the plasma temperature, $kT_{e}$, the plasma optical depth, $\tau$, the geometry switch parameter and the normalization. During the fits, we fixed the redshift $z$ at 0 and the geometry switch parameter to its default value of 1 for a disk geometry. Additionally, we applied the model {\tt tbabs} to account for interstellar absorption along the line of sight, adopting the solar abundance of \citet{wilms00} and the photon ionization cross-section table of \citet{verner96}.

\begin{figure}
\center
\includegraphics[width=0.5\textwidth]{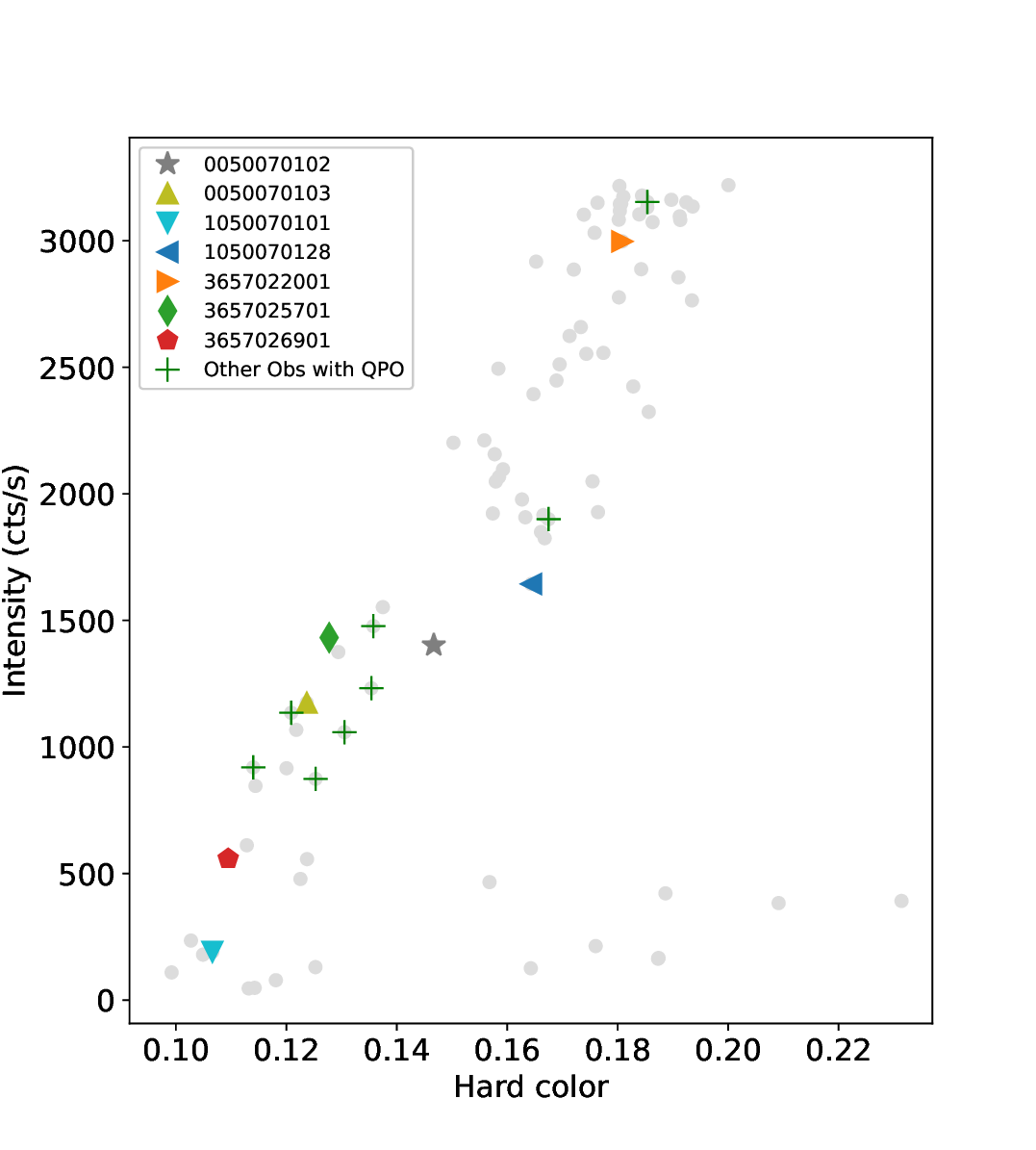}
\caption{
Hardness-Intensity diagram (HID) of 4U 1608--52 using NICER observations. Each gray point in the plot represents a single NICER observation used in this work, and observations with the mHz QPOs are marked with different symbols. Observations with a QPO time less than 1500 s are marked with green crosses. }
\label{HID}
\end{figure}

We then applied the same model to further fit the combined peak, middle, and bottom spectra for each observation. In fitting these spectra, we allowed $kT_{BB}$ and $Nor_{BB}$ to vary freely while fixing all other parameters to the values obtained from the fits to the total spectra. This approach is well-motivated, as the mHz QPOs are proposed to originate from nuclear burning on the neutron star surface; therefore, only the component related to the neutron star radiation should vary across the different mHz QPO stages.

\begin{figure*}
\center
\includegraphics[width=0.45\textwidth]{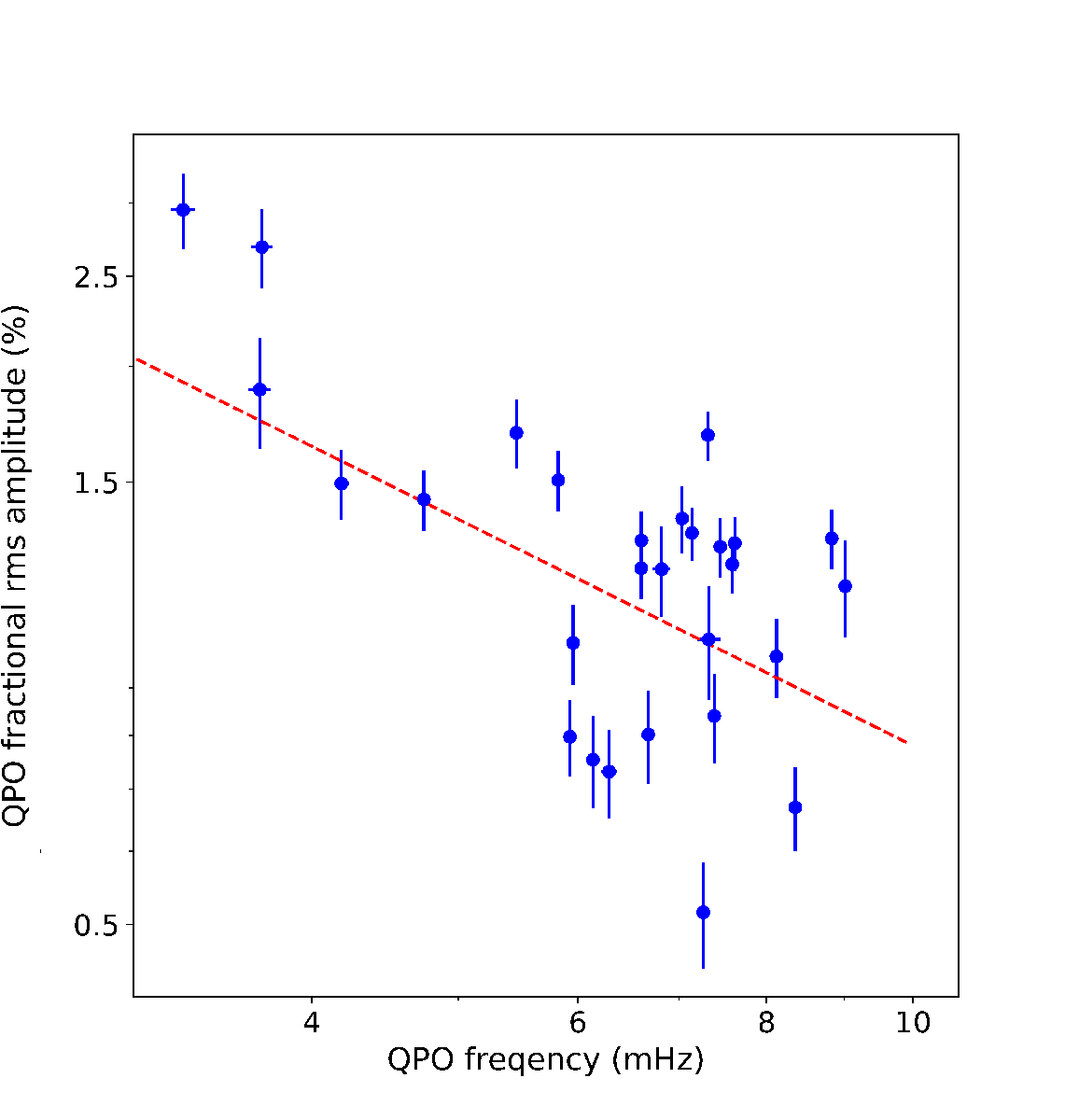}
\includegraphics[width=0.45\textwidth]{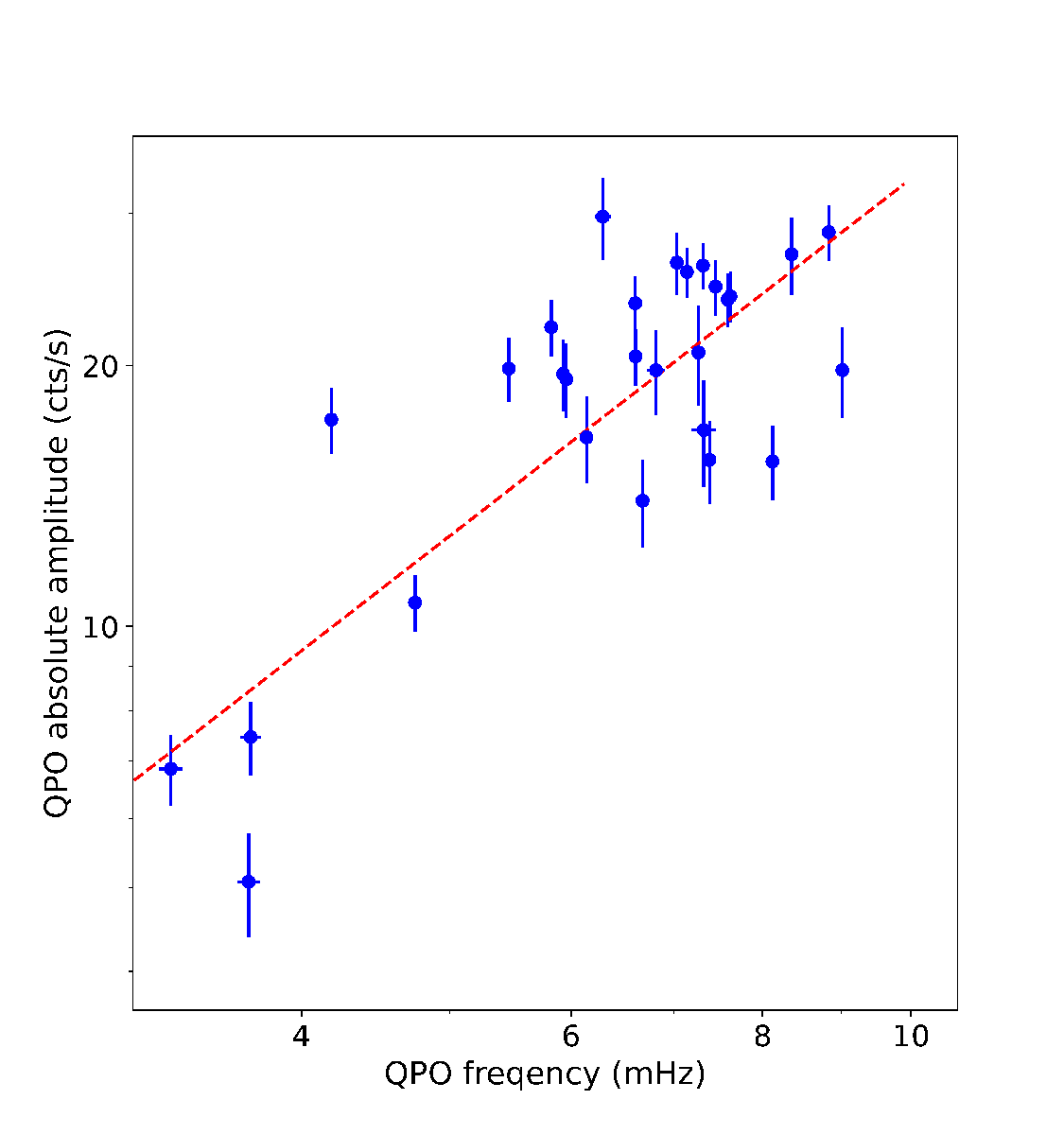}
\caption{QPO fractional rms amplitude (left) and QPO absolute amplitude (right) vs. frequency of the mHz QPOs in 4U 1608--52. The red dashed lines in the plot correspond to the best-fitting power-law model to the data. All errors in the plots are at the 68\% confidence level.}
\label{fre_rms}
\end{figure*}

We further fitted the oscillation component of the QPOs to investigate the spectral properties of the mHz QPOs. We first attempted to fit the S-subtracted S+Q spectra with a {\tt bbodyrad} component; however, the net counts in this case were too low for the spectral analysis. We then fitted the S+Q spectrum and the S spectrum simultaneously using the model {\tt bbodyrad1+comptt+bbodyrad2}, with the first {\tt bbodyrad} component describing the stable flux component from the NS surface and the second {\tt bbodyrad} component fitting the oscillatory flux from the QPO. In the fits, we set the normalization of the {\tt bbodyrad2} to be zero for the S spectra, thereby assuming the oscillation flux is present only in the S+Q spectra. We also linked the parameters of the {\tt bbodyrad1} component and {\tt comptt} component in the S spectra to the corresponding values in the S+Q spectra during the fitting process. In this way, the {\tt bbodyrad2} component in the S+Q spectra describes the spectral component associated with the mHz QPOs. The interstellar absorption parameter $N_{H}$ was fixed at the value obtained from the fits to the total spectra.

\begin{table*}  
\setlength{\tabcolsep}{2pt}
\caption{Best-fitting results for the spectra taken from all the mHz QPO epochs in each NICER observation. All errors in the tables are at the 90 per cent confidence level unless otherwise indicated. A symbol * means that the error pegged at the hard limit of the parameter range.}
\scalebox{0.7}{
\begin{tabular}{ccccccccc}
\hline 
Model Comp        & Parameter    &     0050070102  &     0050070103   &   1050070101  &   1050070128  &    3657022001   & 3657025701   & 3657026901       \\              
\hline                                                                                                                                                                                          
Tbabs   &$N_{\rm H}$ ($10^{22}$ $cm^{-2}$)   &  1.12$\pm$0.02           &  1.16$\pm$0.02           &  1.13$\pm$0.02          &  1.17$\pm$0.02           &  1.12$\pm$0.02               &   1.14$\pm$0.02             &  1.19$\pm$0.03                \\  
\\                                                                                                                                                                                          
Bbodyrad   & $kT_{BB}$ (keV)                 &  0.74$_{-0.07}^{+0.21}$  &  0.65$_{-0.02}^{+0.06}$  &  0.36$\pm$0.02          &  0.81$_{-0.08}^{+0.13}$  &  1.25$_{-0.17}^{+0.20}$      &   0.65$_{-0.01}^{+0.03}$    &  0.43$\pm$0.04       \\                                                                         
           & Norm                            &  207$_{-156}^{+304}$     &  641$_{-344}^{+226}$     &  2478$_{-279}^{+313}$   &  322$_{-152}^{+239}$     &  98$_{-22}^{+59}$            &   910$_{-421}^{+147}$       &  3257$_{-2032}^{+251}$        \\        
\\                                                                             
CompTT     & $kT_{seed}$ (keV)               &  0.41$\pm$0.02           &  0.37$\pm$0.03           &  0.59$_{-0.15}^{+0.12}$ &  0.38$\pm$0.03           &  0.43$\pm$0.01               &   0.39$\pm$0.03             &  0.62$\pm$0.23                \\    
           & $kT_{e}$ (keV)                  &  2.12$_{-0.11}^{+0.15}$  &  2.05$_{-0.05*}^{+0.17}$ &  2.78$_{-0.31}^{+0.90}$ &  2.11$_{-0.11*}^{+0.20}$  &  2.00$_{-0*}^{+0.07}$        &   2.01$_{-0.01*}^{+0.14}$   &  2.15$_{-0.15*}^{+0.43}$      \\                                                                        
           & $\tau$                          &  7.71$\pm$0.78           &  7.93$\pm$0.90           &  6.47$_{-1.33}^{+0.75}$ &  8.69$\pm$0.91           &  8.63$_{-0.60}^{+0.31}$      &   8.37$_{-1.05}^{+0.16}$    &  7.66$_{-1.65}^{+0.83}$       \\                                                          
           & Norm                            &  1.19$\pm$0.05           &  0.93$\pm$0.04           &  0.08$\pm$0.02          &  1.35$_{-0.11}^{+0.07}$  &  2.71$_{-0.20}^{+0.03}$      &   1.07$\pm$0.05             &  0.27$_{-0.08}^{+0.14}$       \\      
           \\                                                                                                                                                                       
     & $\chi_{\nu}^{2}$ ($\chi^{2}/dof$)     & 1.13(1073/952)      &    1.10(1006/917)       &    1.15(900/784)        &   1.02(964/941)        &    0.98(934/952)        &         1.26(1200/952)      &       1.07(904/841)        \\                                                                                                            
                                                                                                                   
\hline                                                                                                                                  
\end{tabular}
}                                                                                 
\medskip  
\label{tot}
\end{table*}

\begin{figure}
\center
\includegraphics[width=0.45\textwidth]{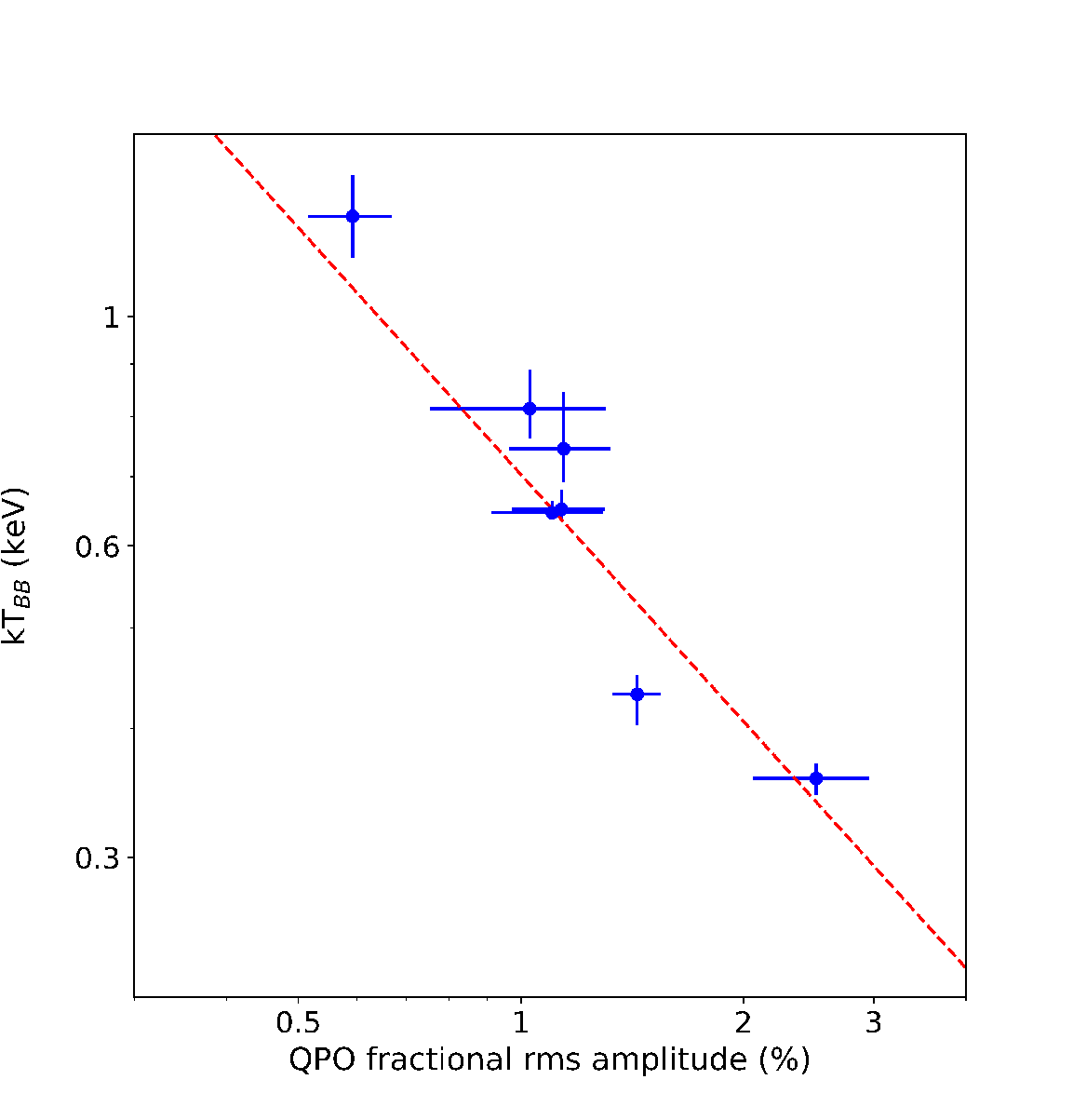}
\caption{Temperature of the blackbody component versus the average fractional rms amplitude of the mHz QPOs in the NICER observations of 4U 1608--52. The red dashed line corresponds to the best-fitting power-law model to the data. The QPO rms amplitude is calculated by averaging the amplitude in different segments in each observation. Here we plot the standard deviation of the rms as the error bar of the amplitude in the case of multiple segments.}
\label{rms-kt}
\end{figure}

\begin{table}
\centering
\caption{Best-fitting results for the mHz QPO intensity-resolved spectra in the NICER observations. }

\scalebox{0.65}{
\begin{tabular}{lccc}
\hline 
Obs.Stage  & $kT_{BB}$ (keV) &  Nor & $\chi_{\nu}^{2}$ ($\chi^{2}/dof$)     \\
\hline     
0050070102.peak    &          0.719$\pm$0.011 &   257$\pm$14   &  1.01 (904/896)  \\     
0050070102.middle  &         0.804$\pm$0.011  &  178$\pm$9     &   1.03 (949/921)      \\     
0050070102.bottom  &        0.698$\pm$0.012   &  212$\pm$13  &   1.15 (1079/935)        \\     
\\
0050070103.peak    &        0.653$\pm$0.006    &  677$\pm$23   & 0.98 (791/809)  \\     
0050070103.middle  &       0.647$\pm$0.007     &  655$\pm$27  & 0.99 (778/783)  \\     
0050070103.bottom  &       0.643$\pm$0.005    &  633$\pm$21   & 1.13 (958/850) \\  
\\   
1050070101.peak    &        0.369$\pm$0.004  & 2226$\pm$102  & 1.04 (578/557)   \\     
1050070101.middle  &       0.361$\pm$0.004   &  2393$\pm$107          & 1.12 (634/566)  \\     
1050070101.bottom  &      0.348$\pm$0.003   &  2681$\pm$96            & 1.06 (679/641)   \\     
\\
1050070128.peak    &      0.825$\pm$0.009   &  332$\pm$13   & 1.04 (901/868)  \\     
1050070128.middle  &     0.807$\pm$0.009   &  332$\pm$14   &  1.00 (869/870)   \\     
1050070128.bottom  &     0.798$\pm$0.010   &  319$\pm$14   &  1.00 (881/881)     \\     
\\
3657022001.peak    &       1.270$\pm$0.016    &  99$\pm$4    &  0.92 (867/941) \\       
3657022001.middle  &       1.220$\pm$0.024     &  100$\pm$7 &  0.96 (848/880) \\       
3657022001.bottom  &      1.192$\pm$0.027     &    95$\pm$7  &  1.01 (878/872)  \\     
\\  
3657025701.peak    &     0.657$\pm$0.004   & 925$\pm$23   & 1.06 (908/855)   \\     
3657025701.middle  &    0.646$\pm$0.005   &  920$\pm$25  &  1.11 (954/859)  \\     
3657025701.bottom  &   0.641$\pm$0.003    &  907$\pm$19  &  1.19 (1109/935)    \\    
\\                                                                                                                                                                              
3657026901.peak    &   0.439$\pm$0.003  &  3179$\pm$94   & 0.89 (587/657) \\             
3657026901.middle  &  0.432$\pm$0.003  &  3251$\pm$79    & 0.95 (697/733) \\             
3657026901.bottom  &  0.427$\pm$0.002  &  3310$\pm$79   & 1.02 (755/740) \\  
                                                                                                                
\hline                                                                                                                                  
\end{tabular}
}
                                                                                 
\medskip  
\label{phase}
\end{table}

\section{Results}
Figure \ref{maxi_swift} shows the long-term light curves of 4U 1608--52 in two different energy bands. We find that nearly all mHz QPOs occur during outbursts, with the only exception being the QPO detected in Obs 1050070101, which lies outside the main outburst phase. In Figure \ref{HID} we show the Hardness-Intensity Diagram (HID) for the NICER observations of 4U 1608--52 analyzed in this work. It is clear that all mHz QPOs appear in the soft and transitional spectral states, while none are detected in the hard state. This finding is consistent with the case of the mHz QPO source 4U 1636--53, for which \citet{lyu19} reported that the mHz QPOs detected in RXTE data were either in the transitional state or the soft spectral state.

In Figure \ref{fre_rms}, we show the correlation between the frequency and the amplitudes of the QPOs. The QPO frequency in different datasets ranges from $\sim$3.3 mHz to $\sim$9.0 mHz, and the fractional rms amplitude is in the range of $\sim$ 0.5\%-2.9\%. As shown in the left panel of Figure \ref{fre_rms}, there is a negative correlation between these two quantities, although the relation is mostly driven by the three data points below 4 mHz. The data points in the plot can be well described by a power-law function $y=a\cdot x^{\beta}$, with the index $\beta = -0.81\pm0.08$. Interestingly, the absolute amplitude (right panel of Figure \ref{fre_rms}) is positively correlated with the QPO frequency, with a power-law index of 1.37$\pm$0.06.

In Table \ref{tot} we present the best-fitting results of the combined spectra of all the mHz QPO epochs for each NICER observation. The temperature of the {\tt bbodyrad} component is distributed in the  $\sim$ 0.36-1.25 keV range, and the electron temperature in the {\tt comptt} component is relatively constant across the observations, at around 2-3 keV. We found that the evolution of the blackbody temperature $kT_{BB}$ agrees well with the classic accretion disk theory: as the source transitions from the soft to the transitional state, the temperature decreases with the declining accretion rate. Furthermore, we found an anti-correlation between the blackbody temperature $kT_{BB}$ and the mHz QPO fractional rms amplitude. As shown in Figure \ref{rms-kt}, $kT_{BB}$ decreases from $\sim$1.2 keV to $\sim$0.4 keV as the rms amplitude of the QPO increases from 0.6\% to 2.5\%. A power-law function could well fit this relation, with the power-law index being $-0.79\pm0.05$. Finally, the fitted {\tt comptt} component indicates that the corona has an optical depth of  $\tau\sim6-9$, with the seed thermal photon temperature in the range of $0.3-0.6$ keV.

\begin{table*} 
\setlength{\tabcolsep}{2pt}
\caption{Best-fitting results for the joint fits of the S and S+Q spectra. }
\scalebox{0.72}{

\begin{tabular}{ccccccccc}
\hline 
Model Comp        & Parameter                &      0050070102             &    0050070103                     &    1050070101                   &      1050070128                 &     3657022001             &      3657025701              &      3657026901                        \\                                                                                                                                                                                                                                                     
\hline                                                                                                                                                                                                                                                                                                                                                                                                                                              
Bbodyrad1   & $kT_{BB}$ (keV)                &    0.82$_{-0.13}^{+0.25}$   &      0.69$_{-0.05}^{+0.12}$       &    0.36$\pm$0.03                &      0.85$_{-0.12}^{+0.19}$     &  1.16$_{-0.13}^{+0.16}$    &  0.68$_{-0.04}^{+0.07}$      &     0.38$_{-0.03}^{+0.12}$              \\                                     
                   & Norm                    &    100$_{-61}^{+163}$       &      340$_{-198}^{+301}$          &    2310$_{-1539}^{+172}$        &      215$_{-99}^{+221}$         &  126$_{-34}^{+59}$         &  457$_{-226}^{+311}$         &     2355$_{-2004}^{+1083}$              \\                                                                                                                                 
\\                                                                                                
CompTT     & $kT_{seed}$ (keV)               &    0.42$\pm$0.01            &      0.38$\pm$ 0.01               &    0.55$_{-0.27}^{+0.19}$       &      0.39$\pm$0.01              &  0.42$\pm$0.01             &  0.41$\pm$0.01               &     0.44$_{-0.08}^{+0.44}$              \\
           & $kT_{e}$ (keV)                  &    2.21$_{-0.13}^{+0.27}$   &         2.11$_{-0.11*}^{+0.26}$   &    2.65$_{-0.40}^{+1.36}$       &      2.15$_{-0.15*}^{+0.33}$    &  2.00$_{-0*}^{+0.12}$      &  2.13$\pm$0.14               &     2.73$_{-0.48}^{+0.91}$              \\                                      
           & $\tau$                          &    7.23$_{-0.81}^{+0.57}$   &        7.28$_{-0.89}^{+0.77}$     &    7.15$_{-2.04}^{+1.24}$       &      8.26$_{-1.17}^{+0.89}$     &  8.65$_{-0.71}^{+0.25}$    &  7.35$_{-0.64}^{+0.73}$      &     5.80$_{-1.22}^{+1.48}$              \\                      
           & Norm                            &    1.15$_{-0.11}^{+0.07}$   &        0.96$\pm$0.08              &    0.08$_{-0.03}^{+0.06}$       &      1.35$_{-0.21}^{+0.12}$     &  2.65$_{-0.27}^{+0.33}$    &  1.12$\pm$0.05               &     0.32$_{-0.20}^{+0.08}$              \\
\\                                                                                                       
Bbodyrad2   & $kT_{BB}$ (keV)                &    0.93$\pm$0.05            &       0.72$\pm$0.07               &    0.67$\pm$0.08                &      1.08$\pm$0.15              &  1.61$_{-0.19}^{+0.22}$    &  0.75$\pm$0.06               &     0.63$\pm0.08$                       \\                                     
                  & Norm                     &    59$\pm$11                &       71$_{-21}^{+28}$            &    38$_{-13}^{+20}$             &      23$_{-9}^{+13}$            &  12$_{-4}^{+6}$            &  73$_{-17}^{+22} $           &     73$_{-26}^{+40}$                    \\                                                                                                                               
                  &  Flux (10$^{-10}$ c.g.s) &   4.78$\pm$0.31   &       2.09$\pm$0.25               &      0.82$\pm$0.11               &   3.44$_{-0.56}^{+0.63}$    & 8.93$_{-1.44}^{+1.71}$ & 2.54$\pm$0.26               &   1.21$\pm$0.18                                      \\
 \\                                                                                                    
       & $\chi_{\nu}^{2}$ ($\chi^{2}/dof$)   &     1.00(1706/1706)     &    1.03(1603/1550)            &     0.99(971/979)           &      1.01(1642/1626)        &  0.99(1706/1729)       &   1.12(1882/1688)        &     0.98(1250/1276)                 \\                                                                                                                                                                                                                                                                                                                                                                                      
\hline                                                                                                                                  
\end{tabular}  

}                                                                               
\medskip  
\label{p_b}
\end{table*}

\begin{table}
\centering
\caption{The column depth of the marginally stable nuclear burning in the NICER observations of 4U 1608--52. We calculated the error at the 68\% confidence level using error propagation method.}
\scalebox{0.85}{
\begin{tabular}{cc}
\hline
 ObsID   & Column depth (10$^{8}$ g cm$^{-2}$)   \\
\hline
0050070102 & 4.6 $\pm$ 0.5  \\
0050070103 & 5.2 $\pm$ 0.6  \\
1050070101 & 11.2 $\pm$ 0.7\\
1050070128 & 4.2 $\pm$ 0.8\\
3657025701 & 5.3 $\pm$ 0.5\\
3657026901 & 8.7 $\pm$ 0.3\\
3657022001 & 3.3 $\pm$ 0.3\\
\hline          
\end{tabular}   
}
\medskip        
\\                  
\label{column}     
\end{table}

\begin{figure*}
\center
\includegraphics[width=0.28\textwidth]{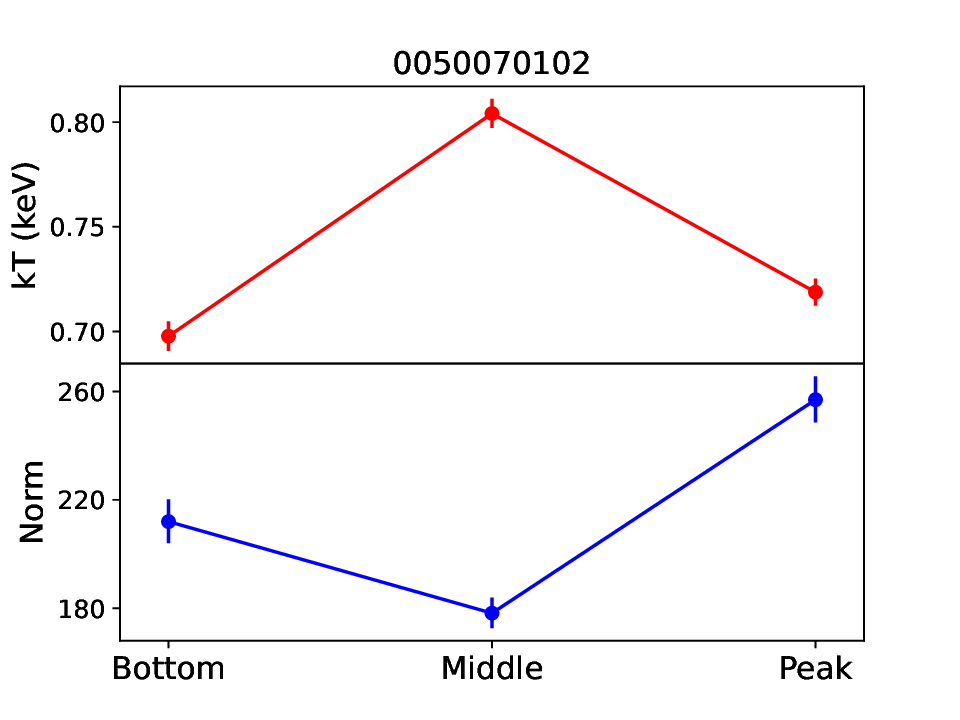}
\includegraphics[width=0.28\textwidth]{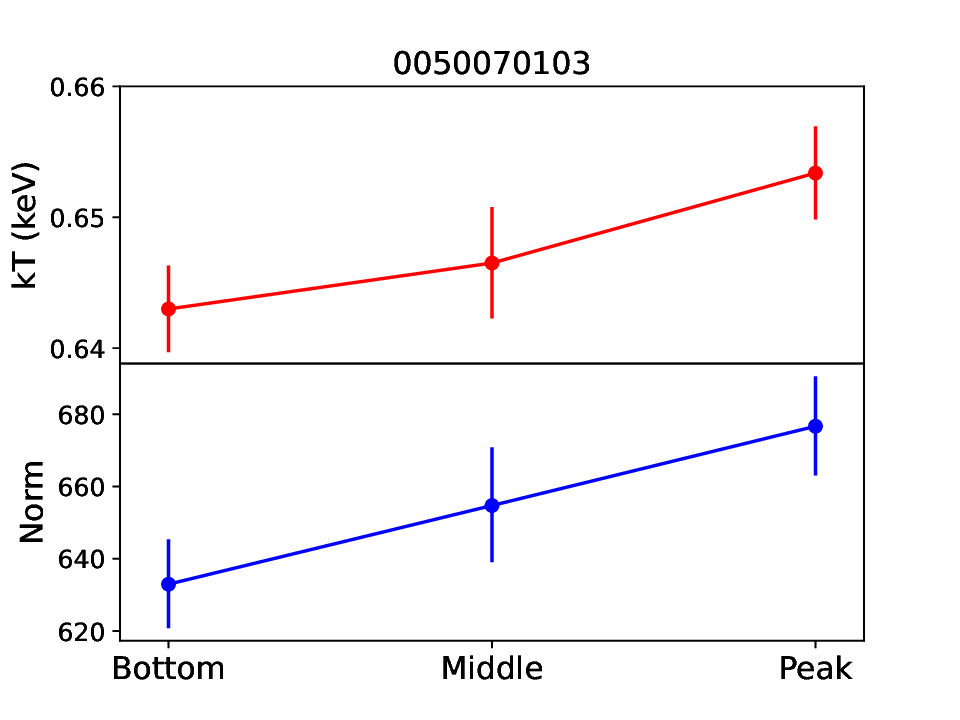}
\includegraphics[width=0.28\textwidth]{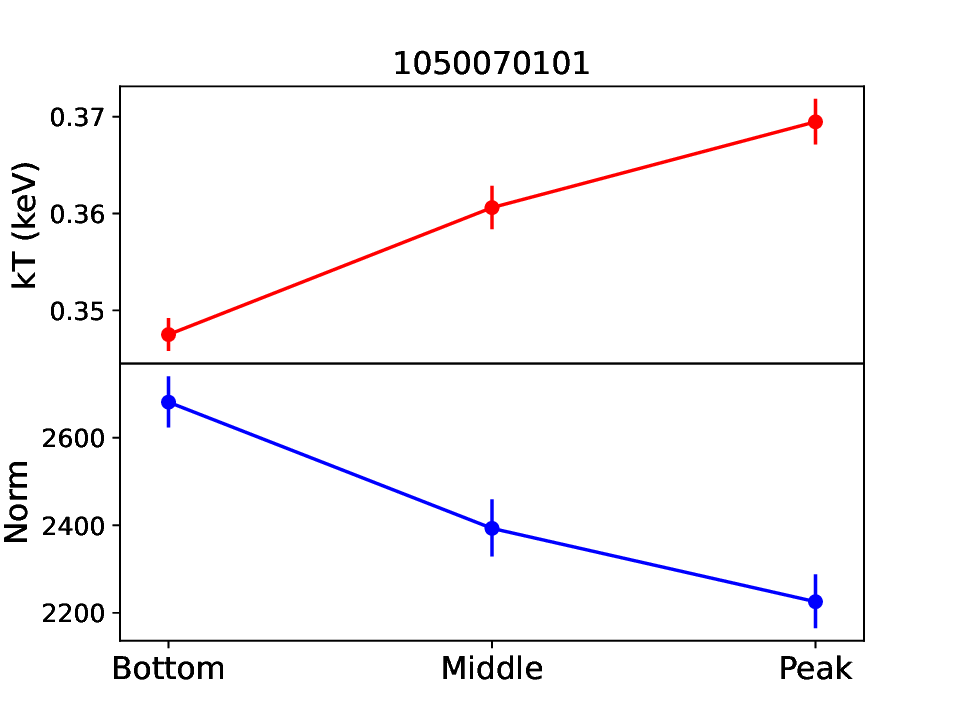}
\includegraphics[width=0.28\textwidth]{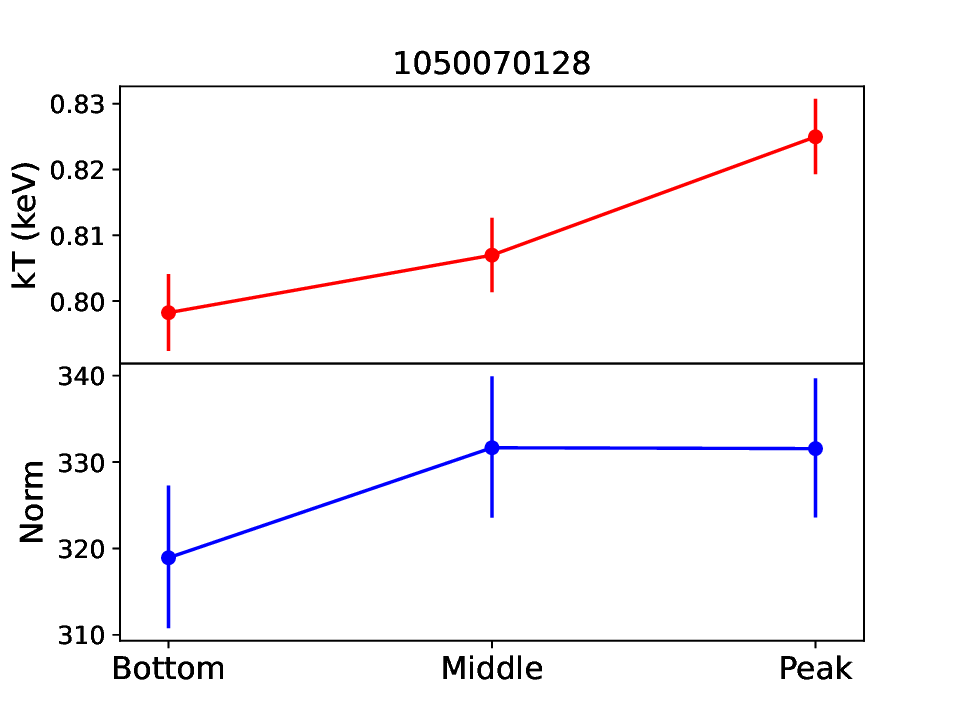}
\includegraphics[width=0.28\textwidth]{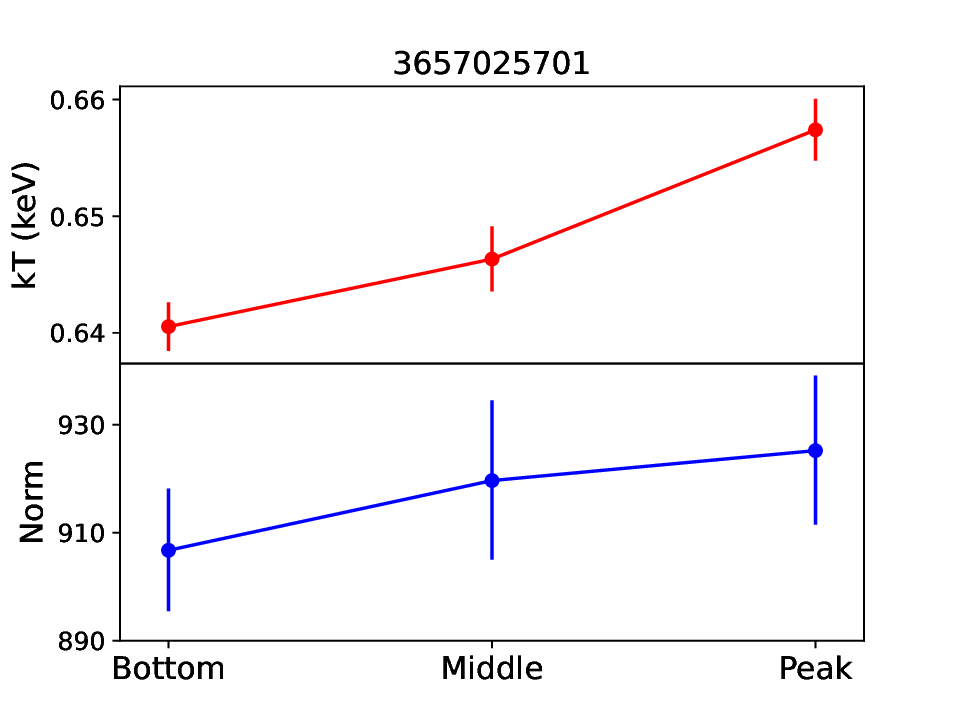}
\includegraphics[width=0.28\textwidth]{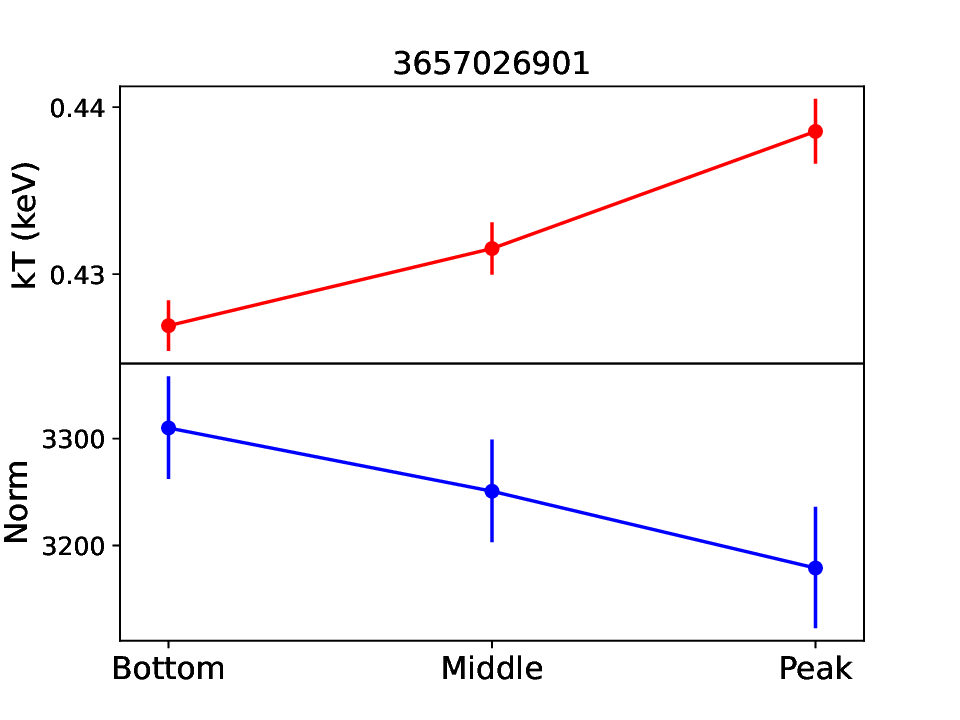}
\includegraphics[width=0.28\textwidth]{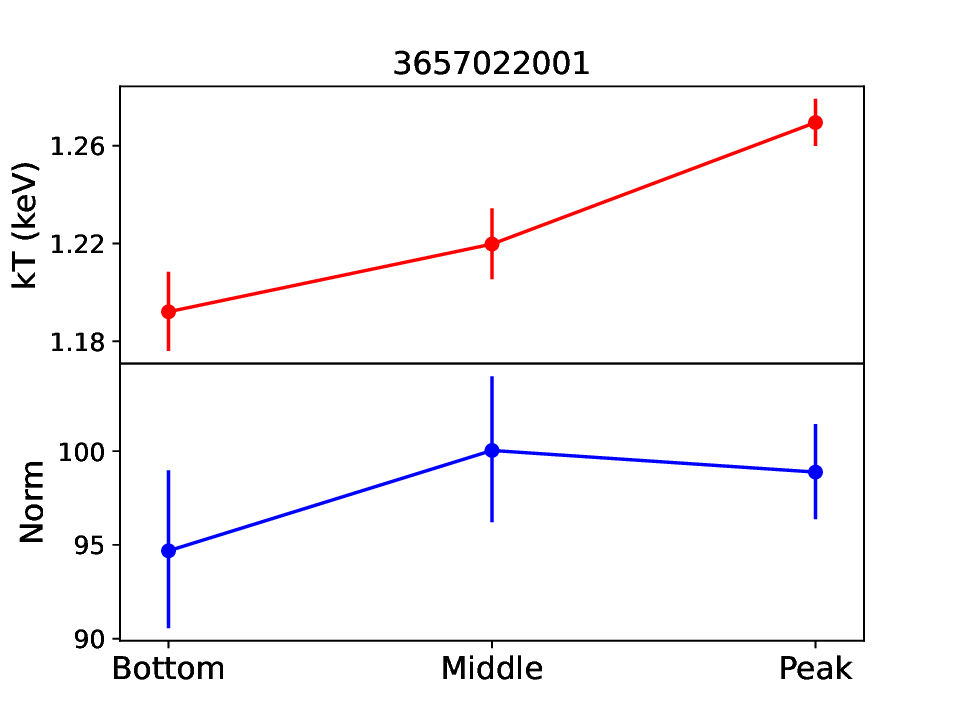}
\caption{Evolution of the blackbody temperature and its normalization across the bottom-middle-peak stage of the mHz QPOs in the NICER observations.}
\label{kTevol}
\end{figure*}

\begin{figure}
\center
\includegraphics[width=0.45\textwidth]{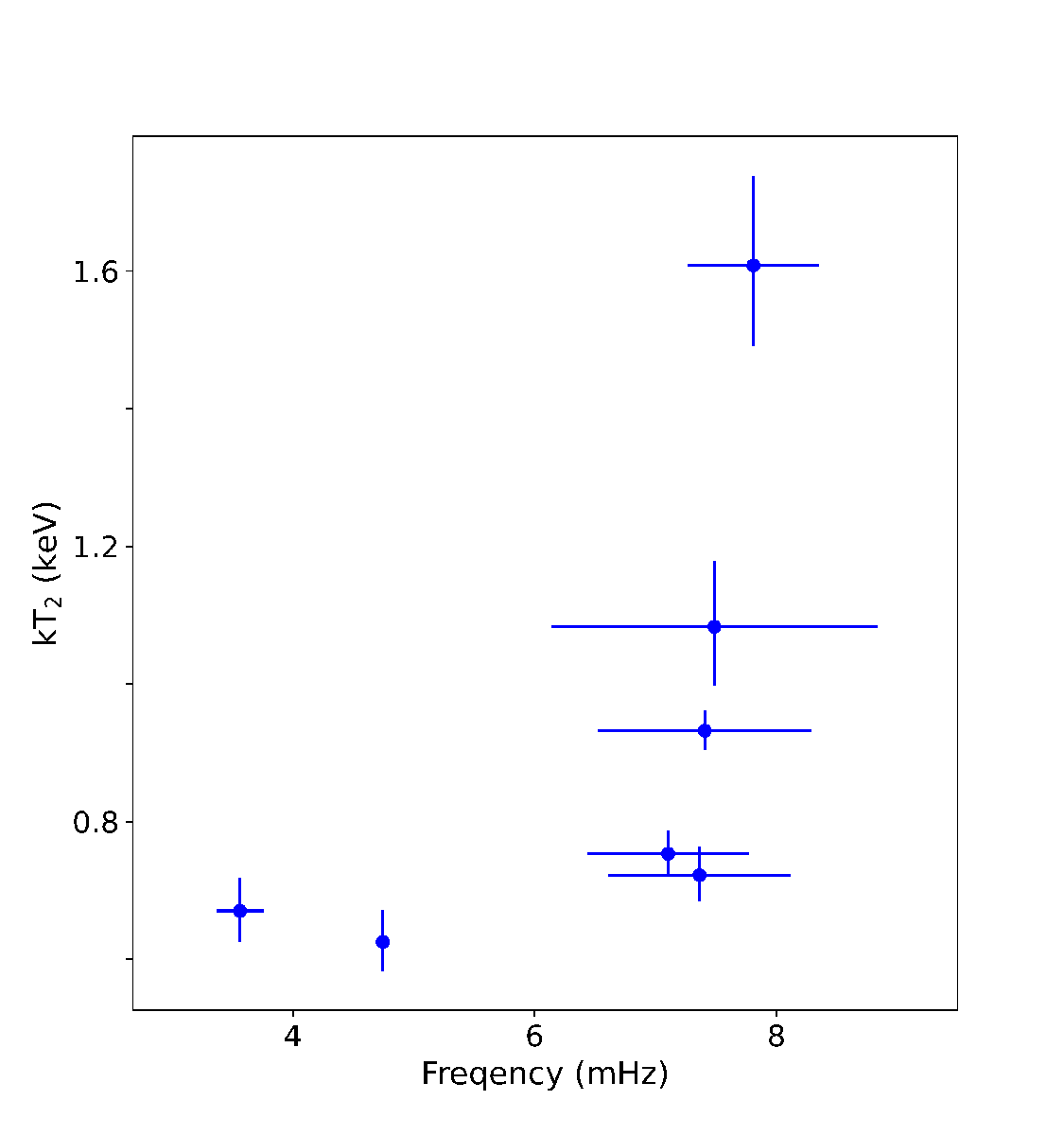}
\caption{Temperature of the 2nd blackbody ({\tt bbodyrad2}) component versus the average frequency of the mHz QPOs in the NICER observations of 4U 1608--52. The average frequency of the QPOs is derived by averaging the QPO frequency in different segments in each observation. We plot the standard deviation of the average frequency as the error bar in the case of multiple segments.}
\label{fre-ktqpo}
\end{figure}

The fitting results of the intensity-resolved spectra of the mHz QPOs are shown in Table \ref{phase}. We show the variation of $kT_{BB}$ and the blackbody normalization across the different stages of the QPO profile in Figure \ref{kTevol}. Regarding $kT_{BB}$, it shows an increasing trend from the QPO bottom stage to the peak stage, except that in observation 0050070102 the temperature suddenly decreases from the middle stage to the peak stage. For the normalization, there is no clear evolutionary pattern in these observations. In observations 1050070101 and 3657026901 the normalization shows a decreasing trend, while in the other observations it remains more or less constant or increases slightly.

In Table \ref{p_b}, we show the best-fitting results for the joint fits of the S and S+Q spectra. We found that the spectra could be well fitted by the model {\tt bbodyrad1+comptt+bbodyrad2}, with reduced $\chi_{\nu}^{2}$ in the range of $\sim$ 0.98-1.12. For each observation, the two blackbody components show a significant difference in the fits: The temperature of the {\tt bbodyrad2} component in the S+Q spectra is higher than that of the {\tt bbodyrad1} component, while its normalization is smaller than that of the {\tt bbodyrad1} component. The unabsorbed flux of the {\tt bbodyrad2} component, which corresponds to the oscillation flux from the mHz QPOs, is $\sim$ $10^{-10}$ erg/s/cm$^{2}$. Finally, as shown in Figure \ref{fre-ktqpo}, the temperature of the {\tt bbodyrad2} component remains roughly constant at frequencies below 6 mHz. At higher frequencies, however, the temperature shows an increasing trend, despite the QPO frequency varying only within a very narrow range.

\section{Discussion}
We studied the timing and spectral properties of the mHz QPOs in 4U 1608--52 with archived NICER observations. We found that all mHz QPOs are present in the soft and transitional spectral state, and no QPO has been detected in the hard state. We also found a positive correlation between the frequency and the absolute amplitude of the QPOs. Additionally, the fractional rms amplitude of the QPO is anti-correlated with both the frequency and the blackbody temperature. Furthermore, the intensity-resolved spectral analysis indicates that the blackbody temperature in most observations shows an increasing trend from the QPO bottom stage to the peak stage. Finally, we found that the radiation from the marginally stable nuclear burning could be well described by an additional blackbody component, {\tt bbodyrad2}. This component shows significant differences from the {\tt bbodyrad1} component, which accounts for the remaining thermal radiation. More importantly, for the first time, we estimated the radiation flux due to marginally stable nuclear burning, which is on the order of 10$^{-10}$ erg s$^{-1}$ cm$^{-2}$.

\subsection{Marginally stable nuclear burning at different depths}
In the mHz QPO epochs, radiation from the marginally stable nuclear burning contributes an additional thermal component to the spectra. Therefore, a second blackbody component is required to explain the spectral properties of the mHz QPOs. \citet{stiele16} applied an additional single blackbody component ({\tt bbodyrad2}) to fit the flux due to the mHz QPOs and derived the temperature and the emitting area related to the QPOs. In their results, the temperature of the blackbody component related to the mHz QPOs is $\sim$0.5-0.8 keV, and the deduced emitting area is below $\sim$ 50 km$^{2}$. The temperature of the other blackbody component ({\tt bbodyrad1}), accounting for the rest thermal radiation from the NS, is much higher, $\sim$ 1.8 keV. Here we also used an additional thermal component but obtained a result which is somewhat different from that of \citet{stiele16}: The temperature of the {\tt bbodyrad2} component is slightly higher than that of the {\tt bbodyrad1} component in each observation, together with a much smaller emitting area in the {\tt bbodyrad2}. Our results suggest that the mHz QPOs originate from a small but hotter region on the neutron star surface, likely located near the equator of the NS as proposed by \citet{lyu16}.

\begin{figure*}
\center
\includegraphics[width=0.9\textwidth]{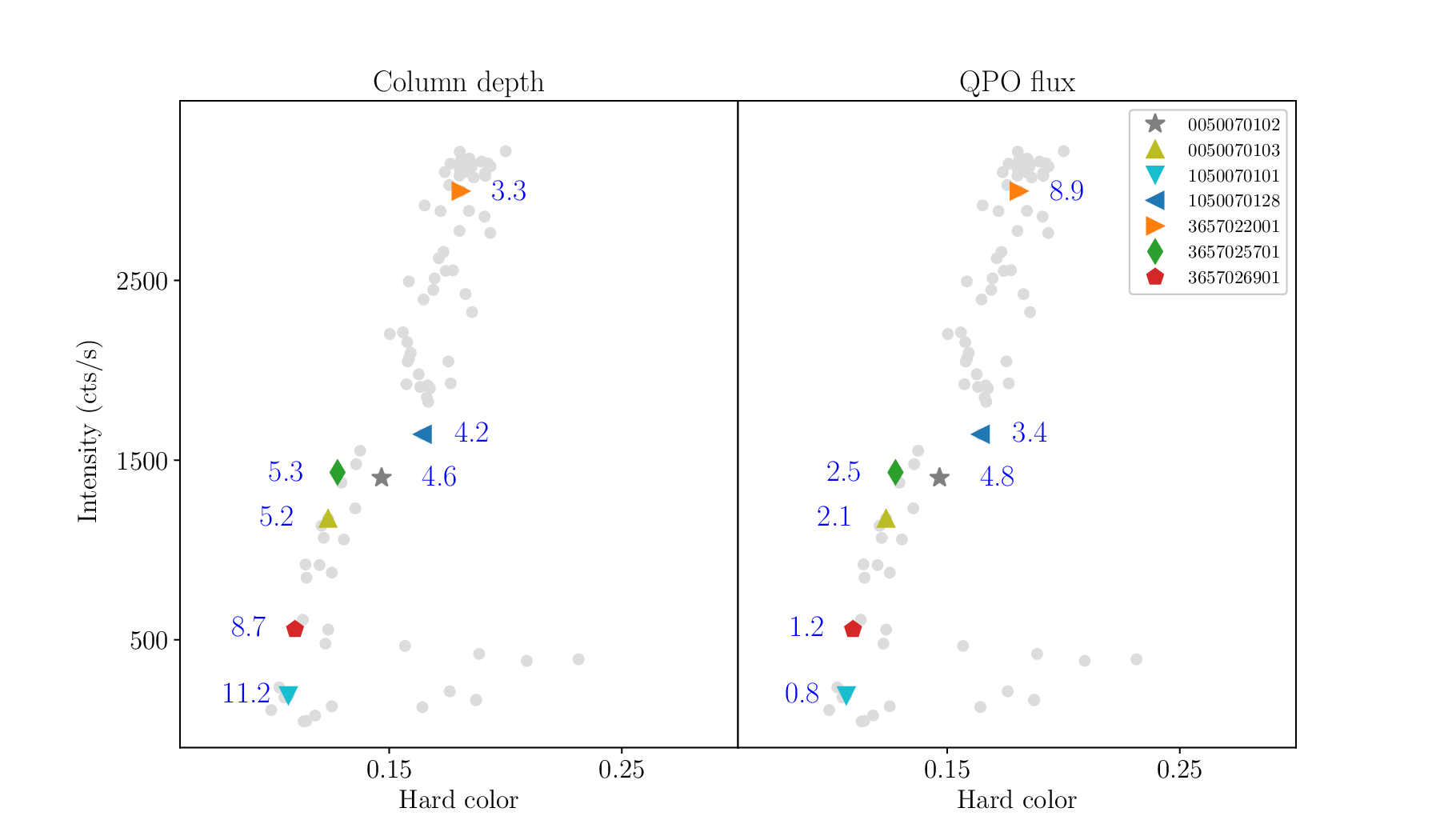}
\caption{Evolution of the column depth of the burning layer $y_{8}$ (10$^{8}$ g cm$^{-2}$) and the oscillation flux (10$^{-10}$ erg s$^{-1}$ cm$^{-2}$) in the NICER observations of 4U 1608--52. We show the values of the column depth and the flux beside the corresponding observations in the plot. The column depth is deduced from the QPO frequency and the temperature of the burning region using equation (\ref{p}). The oscillation flux of the mHz QPOs is calculated from the joint fits of the S+Q and S spectra (Table \ref{p_b}). }
\label{evol_kT}
\end{figure*}

Using the temperature from the {\tt bbodyrad2} component, we can investigate the correlation between the QPO frequency and the temperature of the burning region. Theoretically, the oscillation period is expected to scale with temperature according to the relation \citep{heger07}:
\begin{equation}
P_{\rm osc}=(3.1 \ minutes) \ (\frac{T_{8}}{5})^{1/2} \ y_{8} \ (\frac{\dot{m}}{\dot{m}_{Edd}})^{-1}
\label{p}
\end{equation}
where $T_{8}$ is the temperature in 10$^{8}$ K, $y_{8}$ is the column depth in 10$^{8}$ g cm$^{-2}$, and $\dot{m}$/$\dot{m}_{Edd}$ is the mass accretion rate in Eddington unit. Observationally, \citet{lyu15} investigated the frequency and the average temperature of the NS surface but found no significant correlation between these two quantities. In this work, the relation between the QPO frequency and the temperature of the nuclear burning region (Figure \ref{fre-ktqpo}) is clearly inconsistent with the anti-correlation predicted by Equation (\ref{p}). A possible explanation could be that both the column depth and the accretion rate show variations during the whole burning process. According to \citet{heger07}, $\dot{m}$ refers to the local accretion rate of the burning region, and it should be very close to Eddington when the QPOs are present. Thus, $\dot{m}$/$\dot{m}_{Edd}$ could be treated as a constant during the QPO epochs. This inconsistency between our result and the theoretical prediction therefore indicates that the QPOs in different observations originate from layers with different column depths. 

Assuming $\dot{m}$/$\dot{m}_{Edd}$ $\sim$ 0.925 \citep{heger07}, we further calculated the column depth of the marginally stable nuclear burning using the frequency and the temperature of the {\tt bbodyrad2} component. The derived column depth and its evolution in the HID is shown in Table \ref{column} and Figure \ref{evol_kT}. As shown in the left panel of Figure \ref{evol_kT}, the column depth increases when the source moves from the soft to the transitional state. In the right panel of Figure \ref{evol_kT}, the oscillation flux due to the mHz QPOs shows a decreasing trend as the source evolves to the transitional state. These findings are particular intriguing, as they indicate a link between the marginally stable nuclear burning and the accretion process. Moreover, these observed trends align well with predictions from current marginally stable nuclear burning theory. As shown in Figure 6 in \citet{heger07}, the nuclear energy generation rate shows a decreasing trend as the column depth of the burning layer increases. Therefore, the results in this work support the scenario that, from the soft into the transitional state, nuclear burning occurs at deeper layers and releases less energy, leading to a decreasing QPO flux in observations. 

Why does the depth of the marginally stable nuclear burning increase? We can refer to the theory of unstable burning, which also ignites at deeper layers when the temperature decreases: In a low-temperature environment, more fuel is needed in order to compress the material, heating the layer so that the heating rate exceeding the cooling rate and ultimately igniting the nuclear burning. According to \citet{bildsten98}, the relation between the ignition temperature and the column depth for the unstable burning is $T_{ign} \propto (y_{8})^{-2/5}$. For the marginally stable burning, as the source evolves from the soft to the transitional state, the neutron-star surface becomes cooler, consequently, burning must ignite at deeper layers so that the heating rate approaches the cooling rate. Only under these conditions can the heating process compete with the cooling process and thus generate the quasi-periodic oscillatory flux from the NS surface.

Finally, we can roughly estimate the energy release rate of the marginally stable nuclear burning using the oscillation flux. Assuming that the NS radiation is isotropic, the total power released from the nuclear burning is $P_{ms}=L_{qpo}=4 \pi D^{2}\cdot f_{qpo}$. Taking $D$=3.2 kpc \citep{galloway08} and a QPO flux $f_{qpo}$=$3 \times10^{-10}$ erg s$^{-1}$ cm$^{-2}$ as derived in this work, we obtain $P_{ms}$ $\sim$ 3.7$\times 10^{35}$ erg/s, consistent with the theoretical predictions. Indeed, in the model of \citet{heger07}, the average surface flux due to the mHz oscillations is $\sim$ 0.2$\times10^{24}$ erg s$^{-1}$ cm$^{-2}$ (see Figure 6 in their paper), which corresponds to a total energy release rate of $\sim$ 3.8$\times 10^{35}$ erg/s if the marginally stable nuclear burning covers $\sim$ 15\% of the entire surface of a 10-km NS.

\subsection{Parameter modulations in the mHz QPO period}
There is a long debate about whether the mHz oscillations are due to variations in the blackbody temperature or the emitting area. \citet{stiele16} reported that the mHz QPOs in the two XMM-Newton observations of 4U 1636--53 are caused by  a change in the spatial extent of the emitting region rather than by  a change in the blackbody temperature. Later, \citet{strohmayer18} found that the mHz QPOs in GS 1826--238 are consistent with variations in temperature, with the emitting surface area remaining constant throughout the oscillation cycle. \citet{hsieh20} applied phase-resolved spectroscopy to the mHz QPOs in four XMM-Newton observations of 4U 1636--53. They found that the oscillations are caused by the modulation of the emitting area in three of the four observations, whereas variation in temperature dominates the mHz QPO in the remaining observation. Recently, \citet{xiao25} investigated mHz QPOs in NICER observations of GS 1826--238 and found that the flux oscillation is due to modulation of the temperature of the Comptonization seed photons, which are assumed to originate from the NS surface.

The oscillations in six of the seven observations in this work are due to variations in the blackbody temperature, while in observation 0050070102 neither the temperature nor the normalization shows a clear pattern across the QPO profile (see Figure \ref{kTevol}). Therefore, the mHz QPOs in 4U 1608--52 are mostly caused by the periodic modulation of the burning region temperature, similar to the finding in GS 1826--238. This is understandable, since the thermal radiation flux is proportional to the temperature, $F \propto T^{4}$, so a small variation in temperature can lead to significant modulation of the flux. As for observation 0050070102 in this work, the reason behind this exception remains unknown.

Actually, it remains unclear whether the mHz QPOs in all sources arise from modulation of the same quantity. It is also possible that the dominating factor varies between sources, implying different mechanisms for the flux oscillations. In the future, more observations of mHz QPOs in different systems may help understand the modulation process on the NS surface.

\subsection{Frequency versus amplitude of the mHz QPOs}
Correlations between the typical parameters of the mHz QPOs have been reported in previous work. \citet{lyu15} found no significant correlation between the average frequency of the mHz QPOs and the temperature of the blackbody component in 4U 1636--53 using XMM-Newton plus RXTE data. Later, \citet{lyu19} systematically investigated the timing properties of the mHz QPOs in 4U 1636-53 with RXTE observations and reported no significant correlation between the frequency and the fractional rms amplitude. However, they found a clear correlation between the soft count rate and the fractional rms amplitude. \citet{Mancuso21} reported a potential correlation between the mHz QPO frequency and the average count rate in two cases of 4U 1608--52 using RXTE observations. Here, we found that the frequency of the mHz QPO is anti-correlated with the QPO fractional rms amplitude (see left panel of Figure \ref{fre_rms}), which is inconsistent with the results of \citet{lyu19}. The reason for this discrepancy is unclear. One possibility is that certain properties of mHz QPOs are source-dependent. For example, in 4U 1636--53, the primary parameter driving QPO flux variations is the emitting area \citep{stiele16,hsieh20}, whereas in 4U 1608--52, the results of this work indicate that temperature plays that role. A similar source-dependent behavior may also be present in the frequency–rms relations of these two sources.

The mechanism behind the correlation between the absolute amplitude and the frequency remains unknown. \citet{keek09} found that during the mHz QPO drift stage, as the heat flux from the crust decreases, the frequency decreases by up to several tens of percent, accompanied with an increase in the QPO rms. Since this scenario applies only to the drift stage where the QPO frequency decreases until a X-ray burst occurs, it is unclear whether the relation between the temperature and the amplitude is applicable to our result since we observe no drift behavior in the data. Moreover, the anti-correlation reported by \citet{keek09} is inconsistent with our findings. 

A likely interpretation is that both the frequency and amplitude are controlled by several factors. As shown in Equation (\ref{p}), the oscillation frequency is related to the temperature, column depth and local accretion rate of the burning layer. Regarding the absolute amplitude, related descriptions are very limited in current models. The absolute amplitude of the QPOs should be proportional to the energy released by the marginally stable burning, which is sensitive to the temperature and burning depth. Additionally, the chemical composition of the fuel for the burning also affects the output energy and thus the amplitude. Therefore, the relationship between the frequency and the amplitude is likely not determined by a single factor, but rather by the interplay of these key parameters.

\subsection{How to trigger marginally stable nuclear burning at low accretion rate ? }
It should be noted that, outside outburst period, there is one mHz QPO (ObsID 1050070101) detected, which shows the deepest burning layer and the lowest source intensity. We further calculated the unabsorbed total flux of the source in this observation, $f_{total}$ $\sim$ 1.47 $\times$ 10$^{-9}$ erg s$^{-1}$ cm$^{-2}$ and the corresponding luminosity, $L_{0.1-100\ keV}$ $\sim$ 1.8 $\times$ 10$^{36}$ erg s$^{-1}$. Assuming Eddington luminosity $L_{Edd}$ $\simeq$ 2.5$\times$10$^{38}$ erg s$^{-1}$ \citep{Paradijs94} or $L_{Edd}$ $\simeq$ 3.8$\times$10$^{38}$ erg s$^{-1}$ \citep{kuulkers03}, we found that the mHz QPO occurs at $\lesssim$ 1 \% Eddington limit. This value is about 2 orders of magnitude lower than the accretion rate required to trigger the marginally stable nuclear burning in the current models. 

How can we bridge this wide discrepancy between observations and models? \citet{heger07} proposed that marginally stable burning should be triggered by the local accretion rate at the burning layer. However, it remains unclear whether the local rate can reach the Eddington limit when the global accretion rate is as low as $\sim$1\% of the Eddington rate. Alternatively, \citet{keek09} found that chemical diffusion of the fuel can generate mHz QPOs at around 40\% of the Eddington limit. Combined with additional energy from the crust, rotational mixing, and rotationally induced magnetic fields, this mechanism may help explain the observed transition at 10\% of the Eddington limit. Nevertheless, it is still uncertain whether chemical mixing together with heat from the crust can account for a transition occurring at only $\sim$1\% of the Eddington limit. Further simulations are needed to understand the triggering mechanism of marginally stable nuclear burning at $\lesssim$1\% of the Eddington limit.

\acknowledgments
This research has made use of data obtained from the High Energy Astrophysics Science Archive Research Center (HEASARC), provided by NASA's Goddard Space Flight Center. This research made use of NASA's Astrophysics Data System. Lyu is supported by Hunan Education Department Foundation (grant No. 25B0147). Xiao thanks the support from the Scientiﬁc Research Foundation of Education Bureau of Hunan Province (No. 23A0132)

\bibliographystyle{aasjournal}
\bibliography{paper}

@ARTICLE{revni01,
   author = {{Revnivtsev}, M. and {Churazov}, E. and {Gilfanov}, M. and {Sunyaev}, R.
	},
    title = "{New class of low frequency QPOs: Signature of nuclear burning or accretion disk instabilities?}",
  journal = {\aap},
   eprint = {astro-ph/0011110},
     year = 2001,
    month = jun,
   volume = 372,
    pages = {138-144},
      doi = {10.1051/0004-6361:20010434},
   adsurl = {http://adsabs.harvard.edu/abs/2001A%26A...372..138R}
}

@ARTICLE{heger07,
   author = {{Heger}, A. and {Cumming}, A. and {Woosley}, S.~E.},
    title = "{Millihertz Quasi-periodic Oscillations from Marginally Stable Nuclear Burning on an Accreting Neutron Star}",
  journal = {\apj},
   eprint = {astro-ph/0511292},
     year = 2007,
    month = aug,
   volume = 665,
    pages = {1311-1320},
      doi = {10.1086/517491},
   adsurl = {http://adsabs.harvard.edu/abs/2007ApJ...665.1311H}
}

@ARTICLE{diego08,
   author = {{Altamirano}, D. and {van der Klis}, M. and {Wijnands}, R. and 
	{Cumming}, A.},
    title = "{Millihertz Oscillation Frequency Drift Predicts the Occurrence of Type I X-Ray Bursts}",
  journal = {\apjl},
archivePrefix = "arXiv",
   eprint = {0711.4790},
     year = 2008,
    month = jan,
   volume = 673,
      eid = {L35},
    pages = {L35},
      doi = {10.1086/527355},
   adsurl = {http://adsabs.harvard.edu/abs/2008ApJ...673L..35A}
}

@ARTICLE{linaries12,
   author = {{Linares}, M. and {Altamirano}, D. and {Chakrabarty}, D. and 
	{Cumming}, A. and {Keek}, L.},
    title = "{Millihertz Quasi-periodic Oscillations and Thermonuclear Bursts from Terzan 5: A Showcase of Burning Regimes}",
  journal = {\apj},
archivePrefix = "arXiv",
   eprint = {1111.3978},
 primaryClass = "astro-ph.HE",
     year = 2012,
    month = apr,
   volume = 748,
      eid = {82},
    pages = {82},
      doi = {10.1088/0004-637X/748/2/82},
   adsurl = {http://adsabs.harvard.edu/abs/2012ApJ...748...82L}
}

@ARTICLE{keek09,
   author = {{Keek}, L. and {Langer}, N. and {in't Zand}, J.~J.~M.},
    title = "{The effect of rotation on the stability of nuclear burning in accreting neutron stars}",
  journal = {\aap},
archivePrefix = "arXiv",
   eprint = {0905.4477},
 primaryClass = "astro-ph.HE",
     year = 2009,
    month = aug,
   volume = 502,
    pages = {871-881},
      doi = {10.1051/0004-6361/200911619},
   adsurl = {http://adsabs.harvard.edu/abs/2009A%26A...502..871K}
}

@ARTICLE{keek14,
   author = {{Keek}, L. and {Cyburt}, R.~H. and {Heger}, A.},
    title = "{Reaction Rate and Composition Dependence of the Stability of Thermonuclear Burning on Accreting Neutron Stars}",
  journal = {\apj},
archivePrefix = "arXiv",
   eprint = {1403.4944},
 primaryClass = "astro-ph.HE",
     year = 2014,
    month = jun,
   volume = 787,
      eid = {101},
    pages = {101},
      doi = {10.1088/0004-637X/787/2/101},
   adsurl = {http://adsabs.harvard.edu/abs/2014ApJ...787..101K}
}

@ARTICLE{lyu15,
   author = {{Lyu}, M. and {M{\'e}ndez}, M. and {Zhang}, G. and {Keek}, L.
	},
    title = "{Spectral and timing analysis of the mHz QPOs in the neutron-star low-mass X-ray binary 4U 1636-53}",
  journal = {\mnras},
archivePrefix = "arXiv",
   eprint = {1508.06256},
 primaryClass = "astro-ph.HE",
     year = 2015,
    month = nov,
   volume = 454,
    pages = {541-549},
      doi = {10.1093/mnras/stv1971},
   adsurl = {http://adsabs.harvard.edu/abs/2015MNRAS.454..541L}
}

@ARTICLE{lyu16,
   author = {{Lyu}, M. and {M{\'e}ndez}, M. and {Altamirano}, D. and {Zhang}, G.
	},
    title = "{Millihertz quasi-periodic oscillations in 4U 1636-53 associated with bursts with positive convexity only}",
  journal = {\mnras},
archivePrefix = "arXiv",
   eprint = {1608.06973},
 primaryClass = "astro-ph.HE",
     year = 2016,
    month = dec,
   volume = 463,
    pages = {2358-2362},
      doi = {10.1093/mnras/stw2158},
   adsurl = {http://adsabs.harvard.edu/abs/2016MNRAS.463.2358L}
}

@ARTICLE{lomb76,
   author = {{Lomb}, N.~R.},
    title = "{Least-squares frequency analysis of unequally spaced data}",
  journal = {\apss},
     year = 1976,
    month = feb,
   volume = 39,
    pages = {447-462},
      doi = {10.1007/BF00648343},
   adsurl = {http://adsabs.harvard.edu/abs/1976Ap%26SS..39..447L}
}

@ARTICLE{scargle82,
   author = {{Scargle}, J.~D.},
    title = "{Studies in astronomical time series analysis. II - Statistical aspects of spectral analysis of unevenly spaced data}",
  journal = {\apj},
     year = 1982,
    month = dec,
   volume = 263,
    pages = {835-853},
      doi = {10.1086/160554},
   adsurl = {http://adsabs.harvard.edu/abs/1982ApJ...263..835S}
}

@ARTICLE{stiele16,
   author = {{Stiele}, H. and {Yu}, W. and {Kong}, A.~K.~H.},
    title = "{Millihertz Quasi-periodic Oscillations in 4U 1636-536: Putting Possible Constraints on the Neutron Star Size}",
  journal = {\apj},
     year = 2016,
    month = nov,
   volume = 831,
      eid = {34},
    pages = {34},
      doi = {10.3847/0004-637X/831/1/34},
   adsurl = {http://adsabs.harvard.edu/abs/2016ApJ...831...34S}
}

@ARTICLE{strohmayer18,
   author = {{Strohmayer}, T.~E. and {Gendreau}, K.~C. and {Altamirano}, D. and 
	{Arzoumanian}, Z. and {Bult}, P.~M. and {Chakrabarty}, D. and 
	{Chenevez}, J. and {Guillot}, S. and {Guver}, T. and {Homan}, J. and 
	{Jaisawal}, G.~K. and {Keek}, L. and {Mahmoodifar}, S. and {Miller}, J.~M. and 
	{Ozel}, F.},
    title = "{NICER Discovers mHz Oscillations in the {\ldquo}Clocked{\rdquo} Burster GS 1826{\minus}238}",
  journal = {\apj},
archivePrefix = "arXiv",
   eprint = {1808.04294},
 primaryClass = "astro-ph.HE",
     year = 2018,
    month = sep,
   volume = 865,
      eid = {63},
    pages = {63},
      doi = {10.3847/1538-4357/aada14},
   adsurl = {http://adsabs.harvard.edu/abs/2018ApJ...865...63S}
}

@ARTICLE{Mancuso19,
       author = {{Mancuso}, G.~C. and {Altamirano}, D. and {Garc{\'\i}a}, F. and
         {Lyu}, M. and {M{\'e}ndez}, M. and {Combi}, J.~A. and
         {D{\'\i}az-Trigo}, M. and {in't Zand}, J.~J.~M.},
        title = "{Discovery of millihertz quasi-periodic oscillations in the X-ray binary EXO 0748-676}",
      journal = {\mnras},
         year = "2019",
        month = "Jun",
       volume = {486},
       number = {1},
        pages = {L74-L79},
          doi = {10.1093/mnrasl/slz057},
archivePrefix = {arXiv},
       eprint = {1905.01956},
 primaryClass = {astro-ph.HE},
       adsurl = {https://ui.adsabs.harvard.edu/abs/2019MNRAS.486L..74M}
}

@ARTICLE{lyu20,
       author = {{Lyu}, Ming and {Zhang}, Guobao and {M{\'e}ndez}, Mariano and
         {Altamirano}, D. and {Mancuso}, G.~C. and {Xiang}, Fu-Yuan and
         {Xiao}, Huaping},
        title = "{XMM-Newton and NICER Measurement of the Rms Spectrum of the Millihertz Quasiperiodic Oscillations in the Neutron-star Low-mass X-Ray Binary 4U 1636-53}",
      journal = {\apj},
         year = 2020,
        month = jun,
       volume = {895},
       number = {2},
          eid = {120},
        pages = {120},
          doi = {10.3847/1538-4357/ab8cbe},
archivePrefix = {arXiv},
       eprint = {2006.09563},
 primaryClass = {astro-ph.HE},
       adsurl = {https://ui.adsabs.harvard.edu/abs/2020ApJ...895..120L}
}

@ARTICLE{wilms00,
   author = {{Wilms}, J. and {Allen}, A. and {McCray}, R.},
    title = "{On the Absorption of X-Rays in the Interstellar Medium}",
  journal = {\apj},
   eprint = {astro-ph/0008425},
     year = 2000,
    month = oct,
   volume = 542,
    pages = {914-924},
      doi = {10.1086/317016},
   adsurl = {http://adsabs.harvard.edu/abs/2000ApJ...542..914W}
}

@ARTICLE{verner96,
   author = {{Verner}, D.~A. and {Ferland}, G.~J. and {Korista}, K.~T. and 
	{Yakovlev}, D.~G.},
    title = "{Atomic Data for Astrophysics. II. New Analytic FITS for Photoionization Cross Sections of Atoms and Ions}",
  journal = {\apj},
   eprint = {astro-ph/9601009},
     year = 1996,
    month = jul,
   volume = 465,
    pages = {487},
      doi = {10.1086/177435},
   adsurl = {http://adsabs.harvard.edu/abs/1996ApJ...465..487V}
}

@ARTICLE{hsieh20,
       author = {{Hsieh}, Hung-En and {Chou}, Yi},
        title = "{Phase-resolved Analyses of Millihertz Quasi-periodic Oscillations in 4U 1636-53 using the Hilbert-Huang Transform}",
      journal = {\apj},
         year = 2020,
        month = sep,
       volume = {900},
       number = {2},
          eid = {116},
        pages = {116},
          doi = {10.3847/1538-4357/abacbd},
archivePrefix = {arXiv},
       eprint = {2008.09321},
 primaryClass = {astro-ph.HE},
       adsurl = {https://ui.adsabs.harvard.edu/abs/2020ApJ...900..116H}
}

@ARTICLE{fei21,
       author = {{Fei}, Zhenyan and {Lyu}, Ming and {M{\'e}ndez}, Mariano and {Altamirano}, D. and {Zhang}, Guobao and {Mancuso}, G.~C. and {Xiang}, Fu-Yuan and {Yang}, X.~J.},
        title = "{The Harmonic Component of the Millihertz Quasi-periodic Oscillations in 4U 1636-53}",
      journal = {\apj},
         year = 2021,
        month = dec,
       volume = {922},
       number = {2},
          eid = {119},
        pages = {119},
          doi = {10.3847/1538-4357/ac2501},
archivePrefix = {arXiv},
       eprint = {2109.07686},
 primaryClass = {astro-ph.HE},
       adsurl = {https://ui.adsabs.harvard.edu/abs/2021ApJ...922..119F}
}

@ARTICLE{lyu19,
       author = {{Lyu}, Ming and {M{\'e}ndez}, Mariano and {Altamirano}, D. and {Zhang}, Guobao and {Mancuso}, G.~C.},
        title = "{Discovery of an Accretion-rate Independent Absolute RMS Amplitude of Millihertz Quasi-periodic Oscillations in 4U 1636-53}",
      journal = {\apj},
         year = 2019,
        month = nov,
       volume = {885},
       number = {1},
          eid = {5},
        pages = {5},
          doi = {10.3847/1538-4357/ab44a6},
archivePrefix = {arXiv},
       eprint = {1909.07148},
 primaryClass = {astro-ph.HE},
       adsurl = {https://ui.adsabs.harvard.edu/abs/2019ApJ...885....5L}
}

@ARTICLE{Mancuso21,
       author = {{Mancuso}, G.~C. and {Altamirano}, D. and {M{\'e}ndez}, M. and {Lyu}, M. and {Combi}, J.~A.},
        title = "{Drifts of the marginally stable burning frequency in the X-ray binaries 4U 1608-52 and Aql X-1}",
      journal = {\mnras},
         year = 2021,
        month = apr,
       volume = {502},
       number = {2},
        pages = {1856-1863},
          doi = {10.1093/mnras/stab159},
archivePrefix = {arXiv},
       eprint = {2102.01181},
 primaryClass = {astro-ph.HE},
       adsurl = {https://ui.adsabs.harvard.edu/abs/2021MNRAS.502.1856M}
}

@ARTICLE{xiao25,
       author = {{Xiao}, Hua and {Ji}, Long and {Tsygankov}, Sergey and {Chen}, Yupeng and {Zhang}, Shu and {Li}, Zhaosheng},
        title = "{A Systematic Study of Millihertz Quasiperiodic Oscillations in GS 1826‑238}",
      journal = {\apj},
         year = 2025,
        month = apr,
       volume = {982},
       number = {2},
          eid = {180},
        pages = {180},
          doi = {10.3847/1538-4357/adbcaa},
archivePrefix = {arXiv},
       eprint = {2410.04963},
 primaryClass = {astro-ph.HE},
       adsurl = {https://ui.adsabs.harvard.edu/abs/2025ApJ...982..180X}
}

@ARTICLE{Titarchuk94,
       author = {{Titarchuk}, Lev},
        title = "{Generalized Comptonization Models and Application to the Recent High-Energy Observations}",
      journal = {\apj},
         year = 1994,
        month = oct,
       volume = {434},
        pages = {570},
          doi = {10.1086/174760},
       adsurl = {https://ui.adsabs.harvard.edu/abs/1994ApJ...434..570T}
}

@ARTICLE{galloway08,
       author = {{Galloway}, Duncan K. and {Muno}, Michael P. and {Hartman}, Jacob M. and {Psaltis}, Dimitrios and {Chakrabarty}, Deepto},
        title = "{Thermonuclear (Type I) X-Ray Bursts Observed by the Rossi X-Ray Timing Explorer}",
      journal = {\apjs},
         year = 2008,
        month = dec,
       volume = {179},
       number = {2},
        pages = {360-422},
          doi = {10.1086/592044},
archivePrefix = {arXiv},
       eprint = {astro-ph/0608259},
 primaryClass = {astro-ph},
       adsurl = {https://ui.adsabs.harvard.edu/abs/2008ApJS..179..360G}
}

@INPROCEEDINGS{bildsten98,
       author = {{Bildsten}, L.},
        title = "{Thermonuclear Burning on Rapidly Accreting Neutron Stars}",
    booktitle = {The Many Faces of Neutron Stars.},
         year = 1998,
       editor = {{Buccheri}, R. and {van Paradijs}, J. and {Alpar}, A.},
       series = {NATO Advanced Study Institute (ASI) Series C},
       volume = {515},
        month = jan,
        pages = {419},
          doi = {10.48550/arXiv.astro-ph/9709094},
archivePrefix = {arXiv},
       eprint = {astro-ph/9709094},
 primaryClass = {astro-ph},
       adsurl = {https://ui.adsabs.harvard.edu/abs/1998ASIC..515..419B}
}

@ARTICLE{kuulkers03,
       author = {{Kuulkers}, E. and {den Hartog}, P.~R. and {in't Zand}, J.~J.~M. and {Verbunt}, F.~W.~M. and {Harris}, W.~E. and {Cocchi}, M.},
        title = "{Photospheric radius expansion X-ray bursts as standard candles}",
      journal = {\aap},
         year = 2003,
        month = feb,
       volume = {399},
        pages = {663-680},
          doi = {10.1051/0004-6361:20021781},
archivePrefix = {arXiv},
       eprint = {astro-ph/0212028},
 primaryClass = {astro-ph},
       adsurl = {https://ui.adsabs.harvard.edu/abs/2003A&A...399..663K}
}

@ARTICLE{Paradijs94,
       author = {{van Paradijs}, J. and {McClintock}, J.~E.},
        title = "{Absolute visual magnitudes of low-mass X-ray binaries.}",
      journal = {\aap},
         year = 1994,
        month = oct,
       volume = {290},
        pages = {133-136},
       adsurl = {https://ui.adsabs.harvard.edu/abs/1994A&A...290..133V}
}

\end{document}